\documentclass[twocolumn]{aa}  
\usepackage{amsmath}
\usepackage{xcolor}
\usepackage{yhmath}
\newcommand{\be}{\boldsymbol{e}}
\newcommand{\br}{\boldsymbol{r}}

\newcommand{\buu}{\boldsymbol{u}}
\newcommand{\bw}{\boldsymbol{w}}

\newcommand{\but}{\boldsymbol{\widetilde{u}}}
\newcommand{\bup}{\boldsymbol{u^{\prime\prime}}}

\newcommand{\bnabla}{\boldsymbol{\nabla}}

\newcommand{\bsigma}{\boldsymbol{\sigma}}
\newcommand{\btau}{\boldsymbol{\tau}}
\newcommand{\bB}{\boldsymbol{B}}

\newcommand{\dd}{{\rm d}}

\newcommand\norm[1]{\left\lVert#1\right\rVert}

\usepackage{graphicx}
\usepackage{url}
\usepackage{hyperref}
\hypersetup{
  unicode=true,      
  pdftoolbar=true,    
  pdfmenubar=true,    
  pdffitwindow=true,   
  pdftitle={},
  pdfauthor={Gagnier \& Pejcha},
  pdfkeywords={},     
  pdfnewwindow=true,   
  colorlinks=true,    
  linkcolor=blue,     
  citecolor=blue,     
  filecolor=gray,     
  urlcolor=blue      
}

\usepackage{txfonts}
\begin{document} 
\title{Post-dynamical inspiral phase of common envelope evolution}
\subtitle{The role of magnetic fields}
  \titlerunning{Post-dynamical inspiral of common envelope}
  \authorrunning{Gagnier \& Pejcha} 


\author{Damien Gagnier
\and Ond\v{r}ej Pejcha}

\institute{Institute of Theoretical Physics, Faculty of Mathematics and Physics, Charles University, V Hole\v{s}ovi\v{c}k\'{a}ch 2, Praha 8, 180 00, Czech Republic, \email{damien.gagnier@matfyz.cuni.cz}}

\date{Received 2023}

\abstract{

During common envelope evolution, an initially weak magnetic field may undergo amplification by interacting with spiral density waves and turbulence generated in the stellar envelope by the inspiralling companion. Using 3D magnetohydrodynamical simulations on adaptively refined spherical grids with excised central regions, we studied the amplification of magnetic fields and their effect on the envelope structure, dynamics, and the orbital evolution of the binary during the post-dynamical inspiral phase. About $95\%$ of magnetic energy amplification arises from magnetic field stretching, folding, and winding due to differential rotation and turbulence while compression against magnetic pressure accounts for the remaining $\sim 5\%$. Magnetic energy production peaks at a scale of $3a_\text{b}$, where $a_\text{b}$ is the semimajor axis of the central binary's orbit. Because the magnetic energy production declines at large radial scales, the conditions are not favorable for the formation of magnetically collimated bipolar jet-like outflows unless they are generated on small scales near the individual cores, which we did not resolve. Magnetic fields have a negligible impact on binary orbit evolution, mean kinetic energy, and the disk-like morphology of angular momentum transport, but turbulent Maxwell stress can dominate Reynolds stress when accretion onto the central binary is allowed, leading to an $\alpha$-disk parameter of $\simeq 0.034$.
Finally, we discovered accretion streams arising from the stabilizing effect of the magnetic tension from the toroidal field about the orbital plane, which prevents overdensities from being destroyed by turbulence and enables them to accumulate mass and eventually migrate toward the binary.

}

\maketitle
%
\section{Introduction}\label{sec:intro}


Common envelope evolution (CEE) is a phase in the evolution of binary star systems where one star undergoes significant expansion and engulfs its companion within a shared envelope. The drag experienced by the companion star initiates its rapid spiral-in through the envelope, which leads to the transfer of energy and angular momentum to the envelope gas \citep{Paczynski1976}. CEE leads to two potential outcomes: either the stars merge into a single object or the rapid inspiral slows down. The exact causes of the inspiral deceleration remain elusive, but factors such as reduced gas density or corotation of gas with the companion likely play a major role \citep{Roepke22}. Simulations suggest that the post-dynamical inspiral phase associated with weak drag from the shared envelope can persist for at least hundreds of orbits \citep[e.g.,][]{Ricker2012,Passy2012,Ohlmann2016,Ivanova2016,Gagnier2023}. It is expected that such a weak drag gradually brings the binary closer together and ejects the envelope, ultimately resulting in a post-CEE binary \citep[e.g.,][]{Ivanova2013,Clayton2017,Glanz2018}. 

Accurately modeling CEE presents significant challenges due to the complex interplay of hydrodynamics, magnetohydrodynamics, radiative processes, and stellar evolution. While significant progress has been made through all-encompassing end-to-end simulations, interpreting these simulation results presents significant challenges due to a high degree of complexity resulting from a combination of multiple physical processes.  A complementary approach, focusing on specific phases of CEE or distinct physical processes, is essential to achieve a deeper understanding. For instance,``wind tunnel'' simulations conducted by \cite{MacLeod2015}, \cite{MacLeod2017}, \cite{Cruz-Osorio2020}, and  \cite{De2020} characterized the flow properties around objects embedded within common envelopes to determine drag and accretion coefficients relevant for the rapid inspiral. 

In \cite{Gagnier2023}, we carried out 3D hydrodynamical numerical simulations focusing exclusively on the post-dynamical inspiral phase of CEE. To have control over the simulation and to achieve a longer time step, we excised the central region containing the binary cores and we emulated the preceding dynamical plunge-in by depositing angular momentum in the envelope. This approach enabled us to perform a comprehensive analysis of the envelope structure and evolution, angular momentum transport, the short-term variability of accretion, and the estimation of the orbital contraction timescale. We found that the orbital contraction timescale is long, $\sim 10^3$ to $\gtrsim 10^5$ orbits of the central binary, but it can become shorter than the thermal timescale of many  envelopes. Furthermore, provided that there is gas remaining around the binary, the orbital contraction timescale never reaches zero, because the gas outside of the binary orbit cannot be kept in corotation and it develops spiral waves, which back-react on the orbit. Furthermore, our study of the variability of accretion exhibited remarkable similarities with what is seen in circumbinary disk simulations.

So far, the vast majority of CEE simulations have been performed only in the limit of hydrodynamics. It is possible that additional physical processes may introduce further complexity and potentially impact the results. In particular, in the field of stellar physics, magnetic fields play a central role in governing the dynamics of various systems, including protoplanetary and circumbinary disks. Within these disks, magnetic fields have a significant impact on crucial processes such as angular momentum transport \citep[e.g.,][]{Papaloizou1995,Balbus1998,Sano2004,Ji2006,Pessah2007,Shi2012}, acceleration of winds \citep[e.g.,][]{Bai2013,Lesur2021,Wafflard2023}, and the launching of jets \citep[e.g.,][]{Ferreira1997,Gold2014,Qian2018,Vourellis2019,Saiki2020}. This raises the fundamental question of whether the influence of magnetic fields extends to CEE, where the highly turbulent environment could  be conducive to their amplification. There exists a possibility that magnetic fields could significantly reshape the envelope's structure and dynamics and could impact the binary's orbital evolution. In fact, jet activity has been argued as a fundamental feature of CEE and shaping of planetary nebulae \citep[e.g.,][]{Soker1994,Soker2016,Shiber2019,Hillel2022}.


Due to numerical challenges of 3D magnetohydrodynamic (MHD) simulations of CEE, there are only a select number of papers addressing the potential effects of magnetic fields in this context. The first 3D MHD simulation of CEE was performed by \cite{Ohlmann2016b} who found that the amplification of magnetic energy was insufficient for magnetic fields to become dynamically significant during the dynamical inspiral phase of CEE. \cite{Schneider2019} conducted 3D MHD simulations of the merger of two massive stars, resulting in strong magnetic fields and yielding a rejuvenated merged star appearing younger and bluer than stars of its age. The most recent 3D MHD simulation of CEE was performed by \cite{Ondratschek2022}. Their findings indicate that magnetic fields may have a substantial influence on shaping the morphology of emerging planetary nebulae by launching jet-like outflows from the immediate vicinity of the two central cores. Similar jet-like polar outflows were also observed in the 2D MHD post-common envelope circumbinary disk simulations of \cite{GarciaSegura2021}. This scarcity of research presents an opportunity for conducting more thorough investigations into the role of magnetic fields in CEE.

In this paper, we aim to investigate the amplification of an initially weak seed magnetic field as it interacts with spiral density waves and hydrodynamical turbulence arising in the envelope from the time-changing gravitational force of the central binary system. We examined the influence of the resulting Lorentz force feedback on the structure and dynamics of the envelope and studied the impact on the orbital evolution of the central binary during the post-dynamical phase of CEE.
Building upon our previous work in \cite{Gagnier2023}, we conducted 3D magnetohydrodynamical numerical simulations dedicated to the post-dynamical inspiral phase of the CEE. To emulate the outcome of the preceding dynamical inspiral phase, we injected angular momentum into the envelope following the methodology described in \cite{Morris2006,Morris2007,Morris2009}, \cite{Hirai2021}, and \cite{Gagnier2023}. We excised a central sphere containing the binary and we applied inner boundary conditions controlling the presence or absence of accretion. With these simulations, we performed a detailed analysis of the energy transfer both within and between the kinetic and magnetic energy reservoirs. Subsequently, we used similar techniques and diagnostics as those used by \cite{Gagnier2023} to assess the impact of magnetic fields on the binary separation evolution timescale, the short-term variability of mass accretion onto the binary, the formation of overdensities, and angular momentum transport within the shared envelope.

This work is structured as follows: in Sect.~\ref{sec:model}, we introduce our physical model and describe the numerical setup
used in our common envelope simulations. In Sect.~\ref{sec:results}, we present the results of our simulations. In particular, we study how kinetic and magnetic energy reservoirs are interconnected, and what contributes to their evolution. We then determine the scales at which the energy reservoirs interact using energy transfer analysis. By comparing our results with those of \cite{Gagnier2023}, we assess the impact of magnetic fields on the binary separation evolution timescale, on the short-term variability of mass accretion onto the binary, the formation of overdensities, and on angular momentum transport within the shared envelope. In Sect.~\ref{sec:conclusions}, we summarize and discuss our results.


\section{Physical model and numerical setup}\label{sec:model}

We construct our post-dynamical inspiral model in the inertial frame at rest with the center of mass of the binary. We do not follow the previous evolution of the inspiraling binary, instead, we emulate its outcome following a procedure similar to \cite{Morris2006,Morris2007,Morris2009}, \cite{Hirai2021}, where the envelope is spun-up until a satisfactory amount of total angular momentum is injected. We describe the details of our implementation of this methodology in  \citet{Gagnier2023}.
We set the gravitational constant $G$, the total binary mass $M = M_1 + M_2$, the  primary's  initial radius $R$, and the angular velocity $\sqrt{GM/R^3}$ to unity. The orbital velocity is fixed to $\Omega_\text{orb} = \sqrt{GM/a_\text{b}^3}$, where $a_\text{b} = r_1 + r_2$ is the fixed binary separation,  $M_1$ is the mass of the primary's core located at  $\{r_1,\theta_1,\varphi_1\}$, and $M_2$ is the mass of the secondary object (either a main-sequence star or a compact object) located at $\{r_2,\theta_2,\varphi_2\}$. The orbital period is $P_\text{orb} = 2\pi/\Omega_\text{orb}$. The two objects are not resolved and are considered as point masses. To simplify our model and to make connection with our previous work, we consider an equal mass binary, $q \equiv M_2/M_1 = 1$, on a fixed circular orbit. The mass of the envelope is $M_\text{env} = 2$ in our units. Our choice of initial parameters for the binary and envelope broadly represents results of ab initio simulations for a range of progenitor types, as we detailed in \citet{Gagnier2023}. Because we are most concerned with the angular momentum transport within the common envelope in the two extreme regimes of mass and angular momentum accretion onto the binary rather than the specific details of the individual cores, we excise a central region  encompassing the binary with a radius $r_{\rm in} = 0.625~a_\text{b} = R/10$. In the rest of this Section, we describe the equations used for solving the problem (Sect.~\ref{sec:equations}) and the initial conditions (Sect.~\ref{sec:IC}). We then present our polar averaging implementation (Sect.~\ref{sec:polar}) and the mesh structure (Sect.~\ref{sec:mesh}).

\subsection{Equations of magnetohydrodynamics}\label{sec:equations}

We use Athena++ \citep[][]{Stone2020} to solve the equations of inviscid and ideal magnetohydrodynamics
    \begin{align} \label{eq:mass}
        \frac{\partial \rho}{\partial t} + \bnabla\cdot \rho \buu &= 0\ , \\
        \frac{\partial \rho \buu}{\partial t} + \bnabla \cdot (\rho \buu \otimes \buu - \bB\otimes \bB+ P^\ast \boldsymbol{I}) &= - \rho \bnabla \Phi\ ,\label{eq:mom} \\
        \frac{\partial E}{\partial t} + \bnabla \cdot \left((E+P \boldsymbol{I})\buu - \boldsymbol{B}(\boldsymbol{B} \cdot \buu) \right) &= - \rho \bnabla \Phi \cdot \buu\  , \label{eq:etot}\\
        \frac{\partial \boldsymbol{B}}{\partial t} - \bnabla \times \left( \buu \times \boldsymbol{B} \right) &= 0 \ \label{eq:induc} , 
    \end{align}
where $P^\ast \boldsymbol{I}$ is a diagonal tensor with components $P^\ast = P + B^2/2 = P + P_B$, $E = e +\rho u^2/2 + B^2/2$ , $e$ is the internal energy density, $P = (\Gamma - 1)e $, $\Gamma = 5/3$ is the adiabatic index, and $\Phi(\br)$ is the gravitational potential of the binary, 
\begin{equation}\label{eq:fullphi}
    \Phi(\br) = -\sum_{i=1}^2 \frac{GM_i}{|\br - \br_i|}\ .
\end{equation}
Heaviside–Lorentz units are used for electromagnetic quantities so that the magnetic permeability $\mu_0 = 1$. In Athena++ (and in almost all higher order Godunov codes), the internal energy density is inferred from the difference between the total energy density $E$ and the kinetic and magnetic energy densities. In the vicinity of the inner boundary, the plasma $\beta$ parameter, $\beta = P/P_B$, may  become very small locally, implying that the internal energy density is locally much smaller than the magnetic and kinetic energy densities. As a result, the MHD solver can locally return unphysically small or even negative internal energy densities. We circumvent this issue by solving the internal energy density equation \citep[e.g.,][]{Stone1992,Bryan2014,Takasao2015}
 \begin{equation}\label{eq:eint}
  \frac{\partial \check{e}}{\partial t} + \bnabla \cdot (\buu \check{e}) + (\Gamma - 1)\check{e} \bnabla \cdot \buu = 0 \ ,    
  \end{equation}
  in addition to the total energy density equation Eq.~(\ref{eq:etot}). Eq.~(\ref{eq:eint}) ensures internal energy density positivity. The selection criterion operates on a cell by cell basis as follows

  \begin{align}\label{eq:condition}e = 
      \begin{cases}
           E - \rho u^2/2 - B^2/2, \quad  E - \rho u^2/2 - B^2/2 < (1 - \epsilon_e) e\\
           \check{e}, \quad \text{otherwise} \ .
      \end{cases}
  \end{align}
 We choose $\epsilon_e = 0.05$. The system is completed by the same boundary conditions as in \cite{Gagnier2023} for hydrodynamic variables to model the two regimes of accretion. Either the inner boundary supports
the weight of the primary’s envelope preventing the fluid to flow through it, or the inner boundary is open to angular momentum and mass flow by imposing zero radial gradient of $\rho$, $P$, $u_\theta$, and angular momentum in ghost zones. The horizontal components of the magnetic field are set to zero in inner ghost zones, and are copied from the last radial cell in the domain in outer ghost zones. The component normal to the boundaries is calculated by imposing the divergence-free constraint.

\subsection{Initial conditions}\label{sec:IC}

As in \cite{Gagnier2023}, we assume that the gas in the envelope is initially in hydrostatic equilibrium and that it can be described by a polytropic equation of state ignoring the gas self-gravity and considering purely radial initial profiles. The radial density profiles are
\begin{align}\label{eq:rho_sh}
    \frac{\rho(r)}{\rho(r_{\rm in})} &=  \left( 1 + C \left(\frac{B}{3} \left(\frac{1}{r^3} - \frac{1}{r_{\rm in}^3} \right) - A \left(\frac{1}{r} - \frac{1}{r_{\rm in}} \right) \right) \right)^n,\\
    P(r) &= K \rho^\Gamma \ ,
    \end{align}    
where $n \equiv 1/(\Gamma - 1) = 3/2$ is the polytropic index and
\begin{equation}
\begin{aligned}
   A^\prime &= A \left( \frac{1}{R} - \frac{1}{r_{\rm in}} \right) \ ,    &   B^\prime &= \frac{B}{3} \left ( \frac{1}{R^3} - \frac{1}{r_{\rm in}^3} \right) \ , \\
  C &=   \frac{\kappa  - 1}{B^\prime - A^\prime} \ , &  K &= \frac{(1-\Gamma)(B^\prime - A^\prime)}{\Gamma (\kappa -1) \rho(r_{\rm in})^{1/n}} \ ,\\ \kappa^n &= \rho(R)/\rho(r_{\rm in}) \ .
\end{aligned}
\end{equation}
Density at the inner boundary $\rho(r_{\rm in})$ is calculated from the prescribed total mass of the envelope. The primary star is initially embedded in a low-density medium to which we apply our outer diode-type boundary conditions and in which the envelope will expand later on. To model this low-density medium, we consider an atmosphere in hydrostatic equilibrium with constant ambient sound speed $c_{s, \rm amb}$ assuming $P = \rho c_{s, \rm amb}^2/\Gamma$, which yields $\rho_{\rm ext} \propto \exp \left( \Gamma/c_{s, \rm amb}^2 \left(A/r - B/(3r^3)\right) \right)$. More details on the hydrodynamical initial conditions can be found in \cite{Gagnier2023}. 

\begin{figure} 
\centering
      \includegraphics[width=\columnwidth]{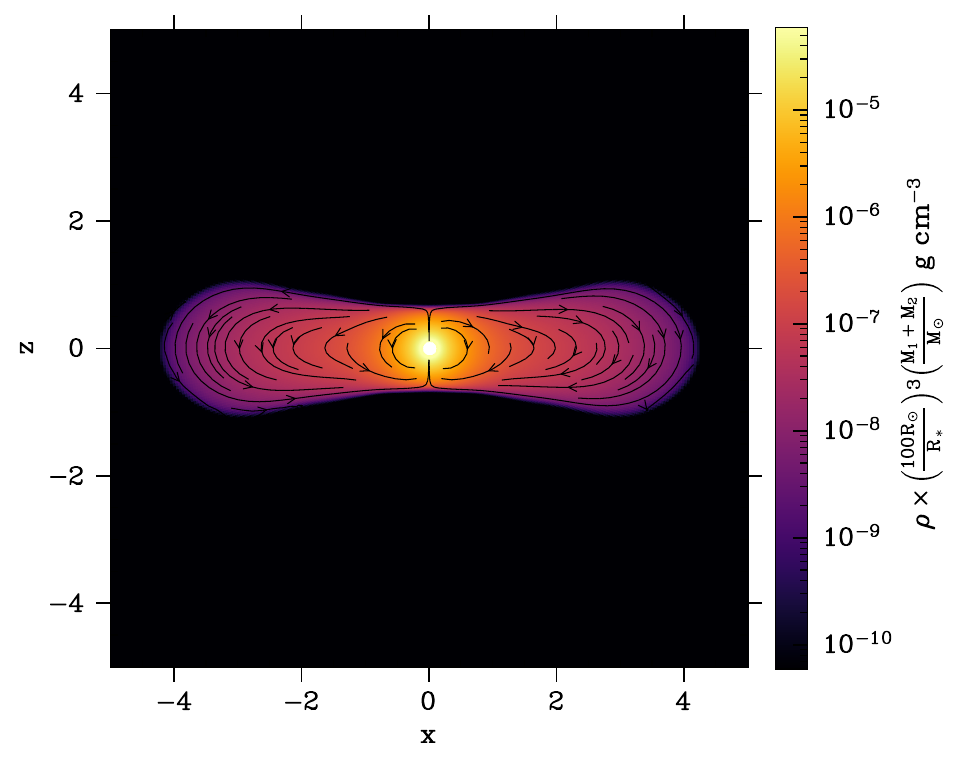} \hfill
   \caption{Zoomed-in snapshot of density cross section  and initial poloidal magnetic field lines in the $xz$ plane after the end of spin-up.}
\label{fig:snapini}
\end{figure}

As is commonly done in (GR)MHD simulations of circumbinary disks \citep[e.g.,][]{Noble2012,Shi2012,Armengol2021}, the magnetic field is initialized as a single poloidal loop within the main body of the envelope at the end of the spin-up phase, with a vector potential $\boldsymbol{A} = \{0,0,A_\varphi\}$ where
\begin{equation}
    A_\varphi = A_0 \max\left(\rho - \rho_\text{cut},0\right) \ ,
\end{equation}
and $\rho_\text{cut} =10^{-5} \rho_\text{max}$. Here, $\rho_\text{max}$ is the maximum density in the envelope at the moment when magnetic field is introduced and $A_0$ is computed to achieve the prescribed initial volume-averaged $\beta$ parameter
\begin{equation}
    \beta_\text{m,i}= \frac{ \int_{\rho > \rho_\text{cut}}  P \dd V }{\int_{\rho > \rho_\text{cut}}  P_B \dd V} \ .
\end{equation}
Specifically, 
\begin{equation}
    A_0 = \left( \frac{ \int_{\rho > \rho_\text{cut}}  2 P \dd V }{\beta_\text{m,i} \int_{\rho > \rho_\text{cut}}  \norm{ \bnabla \times \max \left( \left(\rho - \rho_\text{cut} \right),0 \right) \be_\varphi }^2 \dd V}\right)^{1/2} \ .
\end{equation}
In this work, we consider $\beta_\text{m,i} = 10^3$ in the magnetized region, which corresponds to a field initially too weak to affect the dynamics of the envelope. In Fig.~\ref{fig:snapini}, we present a zoomed-in snapshot of density cross section in the $xz$ plane as well as the poloidal magnetic field lines at the end of spin-up.  In Table~\ref{tab:runs}, we summarize parameters and outcomes of our runs.

\begin{table*}
\caption{Run parameters and simulations outcome at quasi-steady state. \label{tab:runs}}
\begin{center}
\begin{tabular}{lccccccc}
\hline\hline
{Run} &  {$\beta_{m,i}$}  & {Accretion} & {$\alpha_P^\text{QSS}$} & $\alpha_K^\text{QSS}$ & $\alpha_M^\text{QSS}$ & {$R_M^\text{QSS}$} & $P_M^\text{QSS}$\\
\hline
A \citep{Gagnier2023}&  $\infty$ & yes & $0.016$ & $0.034$ & $...$ & ... & $...$ \\
A' \citep{Gagnier2023}&  $\infty$ & no & $-0.026$ & $-0.042$ & $...$ & ... & $...$ \\
B (This work)&  $10^3$ & yes & $0.034$ & $0.043$ & $0.371$ & $939$ & $0.643$ \\
B' (This work)&  $10^3$ & no & $-0.007$ & $-0.020$ & $0.060$ & $899$ & $1.071$  \\
\hline
\end{tabular}
\tablefoot{$R_M^\text{QSS} = VL / \eta_{\rm num}$ is computed using a lengthscale $L$ corresponding to the typical scale of magnetic energy amplification (see Sect.~\ref{sec:trans}), and $V$ is taken as the averaged Alfvén speed in the region where $90\%$ of magnetic energy amplification by line stretching occurs, averaged between $t = 100P_\text{orb}$ and $t = 160P_\text{orb}$. Runs B and B' in this study differ from the ones in \cite{Gagnier2023}.}
\end{center}
\end{table*}

\subsection{Polar averaging}
 \label{sec:polar}
 
The clustering of cells in the azimuthal direction near the polar axis makes the time step given by the Courant–Friedrichs–Lewy condition extremely small. To mitigate this issue, we follow \citet{Gagnier2023} and use a polar averaging technique based on the Ring Average technique \citep{Lyon2004,Zhang2019}. The $r$- and $\theta$-directed face-centered magnetic fields are also averaged in the azimuthal direction near the polar axis and the $\varphi$-directed field is updated from the Faraday's law of induction, accounting for the electric field perturbations resulting from the reconstructed $r$- and $\theta$-directed face-centered magnetic fields \citep[e.g.,][]{Zhang2019}. This procedure ensures $\bnabla \cdot \boldsymbol{B} = 0$ within each cell and each averaged chunk of cells.

At some point in our simulations, the flow ends up stagnating near the polar axis, which leads to the formation of magnetic loops around the axis. If left unattended, the magnetic field can grow to large magnitude. To prevent this, we apply a resistive electric field along the axis after the reconstruction of the face-centered magnetic fields \citep[see also][]{Lyon2004}. The resistive field is given by
\begin{equation}
    \Delta E_r^{\rm axis} \Delta_r= \frac{\cos \theta^\text{axis}}{N_\varphi \tau_{\rm damp}} \sum_{k=1}^{N_\varphi} \Phi^B_\varphi(k) \ ,
\end{equation}
where
\begin{equation}
  \theta^\text{axis}=\begin{cases}
    0 & \text{in the northern hemisphere,} \\
    \pi & \text{in the southern hemisphere.}
  \end{cases}
\end{equation}
Here, $N_\varphi$ is the number of cells in the $\varphi$ direction adjacent to the polar axis at a given radius, $\Delta_r$ is the radial extent of the cell along the polar axis,  $\Phi^B_\varphi(k)$ is the azimuthal magnetic flux through the $k$\textsuperscript{th} cell face, and $\tau_{\rm damp}$ is an arbitrary damping timescale which we choose to be a few time steps, $\mathcal{O}(10^{-5}P_\text{orb})$. By construction, the application of such a resistive electric field leaves $\bnabla \cdot \boldsymbol{B}$ unchanged. While primarily serving magnetic field stabilization, the application of such a resistive electric field may also result in the suppression of jets.

\subsection{Mesh structure and convergence}
\label{sec:mesh}

At the root level, we set $512~\times~256~\times~256$ active cells in $\{r,\theta,\varphi\}$. The radial domain extends from $r = 0.1$ to $r = 10$ with geometric grid spacing so that the aspect ratio of the cells is approximately constant over the entire range of radial scales. The grid spacing in $\theta$ and $\varphi$ directions is uniform with $0 < \theta < \pi$ and $0 < \varphi < 2\pi$. We use polar boundary conditions in the $\theta$ direction allowing free-flow through the pole \citep[][]{Zhu2018,Stone2020}, and periodic boundary conditions in the $\varphi$ direction.  On top of the root level, we add one level of adaptive refinement with criteria based on the azimuthal average of the second derivative error norm $\chi$ of a function $\sigma$ \citep[e.g.,][]{Lohner1987,PLUTO2012}. The norm $\chi$ is defined as
\begin{equation}\label{eq:Lohner}
 \chi = \sqrt{   \frac{\sum_d |\Delta_{d,+1/2} \sigma  - \Delta_{d,-1/2} \sigma|^2                }{\sum_d \left(|\Delta_{d,+1/2} \sigma |  + |\Delta_{d,-1/2}| + \epsilon \sigma_{d,\rm ref}  \right)^2}} \ge \chi_r\ .
\end{equation}
Here, $\Delta_{r,\pm 1/2} = \pm (\sigma_{i \pm 1} - \sigma_i) $ and $\sigma_{r,\rm ref} =  |\sigma_{i+1}| + 2 |\sigma_i| + |\sigma_{i-1}|$. The value of the threshold $\chi_r$ depends on the specifics of the problem and on the chosen refinement variable $\sigma$. Here, we take $\sigma = |\boldsymbol{B}|$.  

\begin{figure} 
\centering
      \includegraphics[ width=88mm]{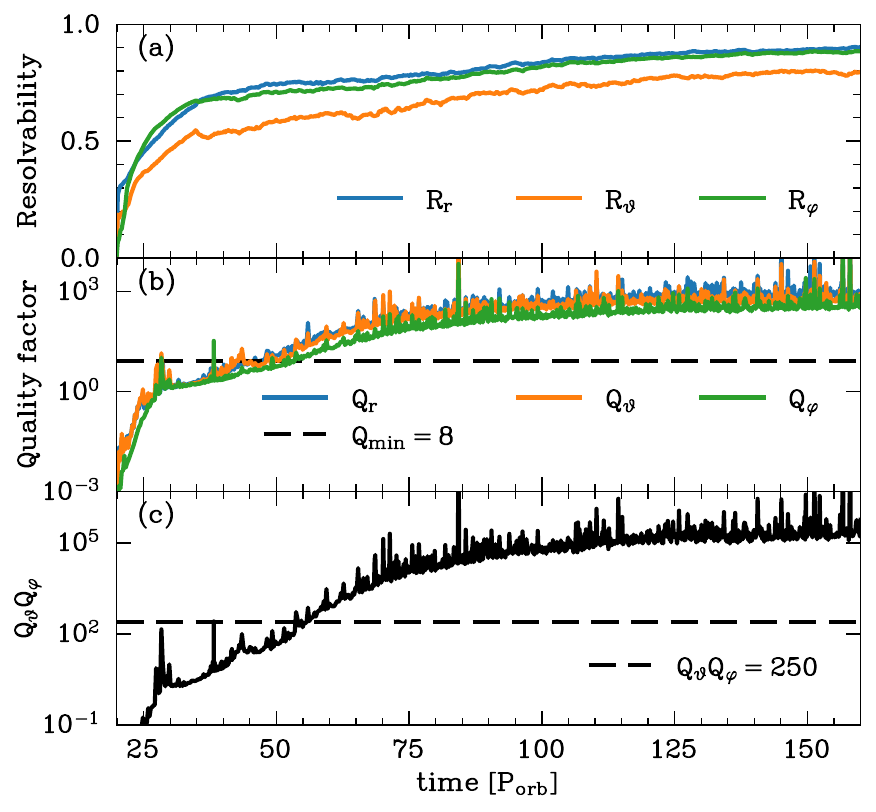} \hfill
   \caption{Temporal evolution of resolvability (panel a), quality factor (panel b), and $Q_\theta Q_\varphi$ (panel c) in simulation run B. As the envelope expands within the numerical domain, these quantities progressively reach values that exceed the threshold values commonly associated with numerical convergence (horizontal dashed lines).}
\label{fig:res}
\end{figure}

We ensure that magnetorotational instability (MRI), which could occur in the simulation, is well resolved. We compute the quality factors in the three directions \citep[e.g.,][]{Noble2010,Hawley2013} defined as
\begin{equation}
    Q_i = \frac{\lambda_\text{MRI,i}}{\Delta_i} =  \frac{2 \pi |v_{A,i}|}{\Omega \Delta_i} \ ,
\end{equation}
where $v_{A,i} = \sqrt{B_i^2/\rho} $ is the $i$-component of the Alfvén speed in code units, and $\Delta_i$ is the size of individual cells in the $i$-direction. Using $Q_i$, we measure the resolvability $R_i$, which is the fraction of cells with a quality factor larger than that associated with marginal resolvability $Q_{\rm min} = 8$  in the direction $i$ \citep[e.g.,][]{Sorathia2012,Parkin2013}. In Figure~\ref{fig:res}, we show evolution of resolvability, volume-averaged quality factor, and the product $Q_\theta Q_\varphi$, for simulation run B. A value exceeding $200$--$250$ for the product $Q_\theta Q_\varphi$ is commonly regarded as a reliable indicator of convergence in MRI simulations \citep[e.g.,][]{Narayan2012, Porth2019, Dhang2023}. We find that while initially small due to limited envelope extent within the numerical domain, these quantities increase  as the envelope expands and reaches the outer boundary, and surpass the threshold for convergence at $t \simeq 55P_{\rm orb}$. This behavior can be attributed to the zero magnetic field amplitude we initially set in the low-density ambient medium encompassing the envelope. We conclude that our simulations have achieved numerical convergence.

\section{Results}\label{sec:results}

\begin{figure*} 
\centering
      \includegraphics[width=180mm]{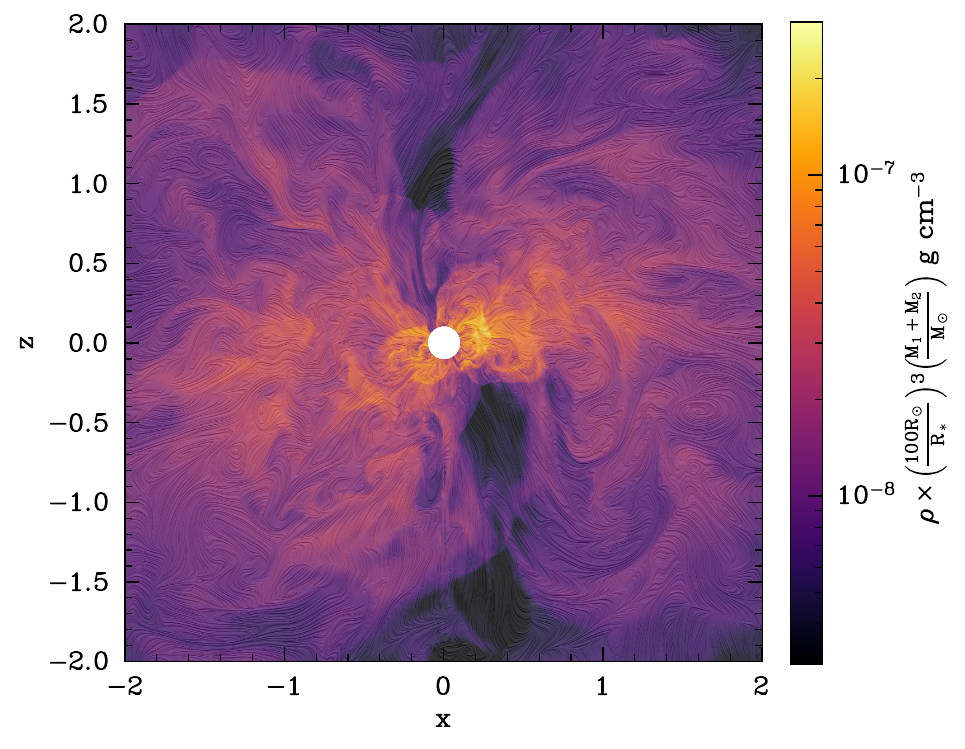} \hfill
   \caption{Zoomed-in snapshot of density cross section in the $xz$ plane at $t=140P_\text{orb}$ for simulation B. The texture, computed using the publicly available Line Integral Convolution Knit python package \citep{LICK}, indicates the meridional streamlines.}
\label{fig:snap925}
\end{figure*}
\begin{figure*} 
\centering
      \includegraphics[width=180mm]{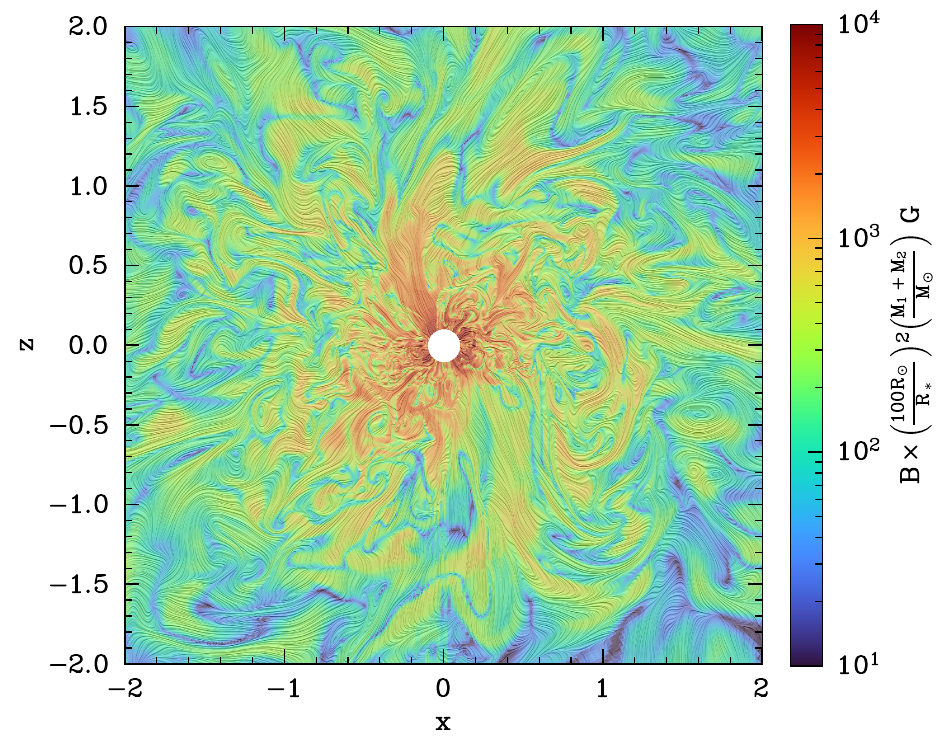} \hfill
   \caption{Zoomed-in snapshot of magnetic field amplitude cross section at $t=140P_\text{orb}$ for simulation B. The texture indicates meridional magnetic field lines.}
\label{fig:snap925_B}
\end{figure*}

We use a total of 4 million CPU hours on the Karolina cluster at IT4Innovations to perform our simulations. The parameters used for our simulation runs are outlined in Table~\ref{tab:runs}. In Figs.~\ref{fig:snap925} and \ref{fig:snap925_B}, we present a zoomed-in snapshot of density and magnetic field amplitude cross section in the $xz$ plane at $t=140P_\text{orb}$ for simulation B. In the rest of this section, we investigate the approach to quasi-steady state (Sect.~\ref{sec:QSS}), kinetic and magnetic energy budget (Sect.~\ref{sec:energy_budget}), evolution of the binary orbit (Sect.~\ref{sec:binary_evolution}), angular momentum transport (Sect.~\ref{sec:AMtransport}), value of the $\alpha$ disk parameter (Sect.~\ref{sec:alpha}), time variability of accretion (Sect.~\ref{sec:accretion_variability}), and behavior of the lump (Sect.~\ref{sec:lump}).

\subsection{Quasi-steady state}\label{sec:QSS}

After an initial transient period, the envelope settles into a quasi-steady state (QSS) during which the global properties of the envelope's dynamics change only slowly. During this phase, matter continues to slowly accrete onto the central binary (if accretion is allowed) and simultaneously escapes from the numerical domain by the large scale flow. To asses the turbulent fluxes of angular momentum and their quasi-steadiness, we apply Reynolds decomposition to the density, the gravitational potential of the binary, and the magnetic field
\begin{align}
\rho &= \overline{\rho} + \rho^\prime \ , \\
\Phi &= \overline{\Phi} + \Phi^\prime \ , \\
B_i &= \overline{B_i} + B_i^\prime \ ,
\end{align}
where overlines indicate azimuthal averages of quantity $q$,
\begin{equation}
    \overline{q} = \frac{1}{2 \pi}\int_0^{2\pi} q \dd \varphi \ .
\end{equation}
We also apply Favre decomposition to the velocity field \citep{Favre1965,Favre1969}
\begin{equation}
    u_i = \widetilde{u_i} + u_i^{\prime\prime} \ ,
\end{equation}
where $\widetilde{q} = \overline{\rho q}/\overline{\rho}$ is the density-weighted Reynolds average also known as Favre average. Taking the cross product of $\br$ with the momentum equation~(\ref{eq:mom}), multiplying by $\be_z$ and applying Reynolds average to the result yields the Reynolds averaged angular momentum evolution equation,
\begin{equation}\label{eq:AMcons}
\begin{aligned}
    \frac{\partial \overline{\rho} s \widetilde{u_\varphi}}{\partial t}  =  - \bnabla \cdot \left( s \left(\mathcal{W}_{r\varphi} + \mathcal{W}_{\theta\varphi} \right) \right) - \overline{\rho s \frac{\partial \Phi^\prime}{\partial \varphi}} \ ,    
\end{aligned}
\end{equation}
where $s = r \sin \theta$ is the radial cylindrical coordinate,
\begin{equation}\label{eq:Wip}
\mathcal{W}_{i \varphi} =  W_{i \varphi} +  w_{i \varphi} \\
\end{equation}
is the total stress and
\begin{align}\label{eq:comp1}
W_{i \varphi} & = \overline{\rho}\,\widetilde{u_i}\,\widetilde{u_\varphi}  - \overline{B_i}\,\overline{B_\varphi}\ , \\ w_{i \varphi} &= \overline{\rho} \, \widetilde{u_i^{\prime\prime} u_\varphi^{\prime\prime} }  - \overline{B_i^\prime B_\varphi^\prime}\ .\label{eq:comp2}\end{align}

\begin{figure} 
\centering
    \includegraphics[ width=88mm]{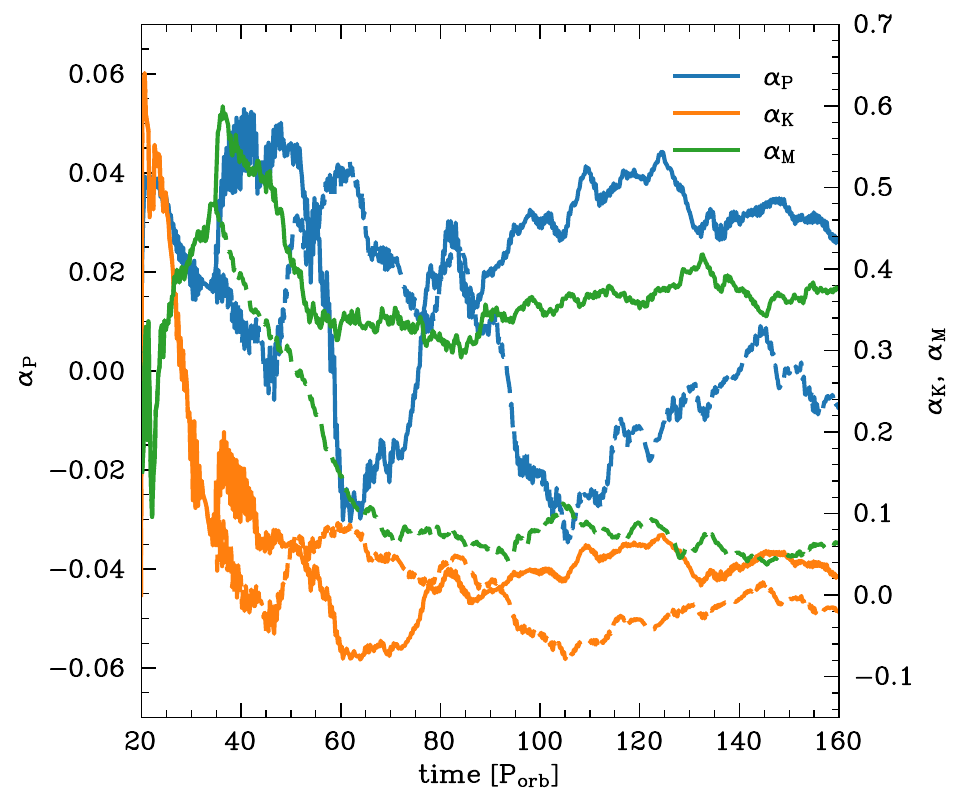} \hfill
   \caption{Evolution of the normalized volume-averaged stresses for simulation runs B (full lines) and B' (dashed lines). We consider a quasi-steady state to be reached after $t \simeq 100P_\text{orb}$. }
\label{fig:alpha_p}
\end{figure}

We assume that a quasi-steady state is reached when the volume-averaged turbulent stress normalized to gas pressure \citep{SS73},
\begin{equation}\label{eq:alphap}
 \alpha_P =\frac{\langle w_{r\varphi} \rangle }{\langle P \rangle}  \ ,
\end{equation}
and the  normalized $r\varphi$--component of the Reynolds and Maxwell stresses, respectively
\begin{align}\label{eq:alphaK}
 \alpha_K &=\frac{\langle 2 \rho  u_r^{\prime\prime} u_\varphi^{\prime\prime} \rangle }{\langle\rho |\boldsymbol{u^{\prime\prime}}|^2\rangle} \ ,  \\
 \alpha_M &=-\frac{\langle 2B_r^\prime B_\varphi^\prime \rangle }{\langle |\boldsymbol{B^\prime}|^2\rangle} \ , \label{eq:alphaM}
\end{align}
become statistically time-independent \citep[e.g.,][]{Simon2012,Parkin2013}. We show the time evolution of $\alpha_P$, $\alpha_K$ and $\alpha_M$ in Fig.~\ref{fig:alpha_p} for runs B and B'. We find that these three quantities reach quasi-steady values from $t \simeq 100P_\text{orb}$ for simulation run B. $\alpha_K$ and $\alpha_M$ also reach quasi-steady values from $t \simeq 100P_\text{orb}$ for simulation run B', however, as $\alpha_K$ is small, it can change sign and lead to a substantial relative variation in $\alpha_P$. Still, we assume that the system has reached a quasi-steady state from $t \simeq 100P_\text{orb}$ for the two simulation runs. The time-averaged values of $\alpha_K$, $\alpha_P$ and $\alpha_M$ at QSS are summarized in Table~\ref{tab:runs}.
The values of $\alpha_P$ and $\alpha_M$ at QSS for run B are in agreement with global and local  magnetohydrodynamic simulations of accretion disks \citep[e.g.,][]{Simon2012,Hawley2013,Parkin2013} and indicate  globally outward turbulent transport of angular momentum. These results, however, do not imply that the angular momentum turbulent flux cannot be locally directed inward nor that Maxwell stress dominates Reynolds stress. The radial and latitudinal dependence of angular momentum radial transport is investigated in Sect.~\ref{sec:AMtransport}. 

\subsection{Energy budget and dynamical relevance of magnetic fields}
\label{sec:energy_budget}

In this section, we determine how kinetic and magnetic energy reservoirs are interconnected, what contributes to their evolution, and the dynamical relevance of magnetic fields during the post-dynamical phase of CEE. In particular, we identify the source of magnetic energy amplification in Sects.~\ref{sec:ME} and \ref{sec:KE}, and determine the scales on which energy is transferred within and between energy reservoirs in Sect.~\ref{sec:trans}. For the sake of brevity, most of this analysis is restricted to simulation run B, but results for B' are analogous.

\subsubsection{Magnetic energy budget}\label{sec:ME}
\begin{figure} 
 \centering
       \includegraphics[ width=88mm]{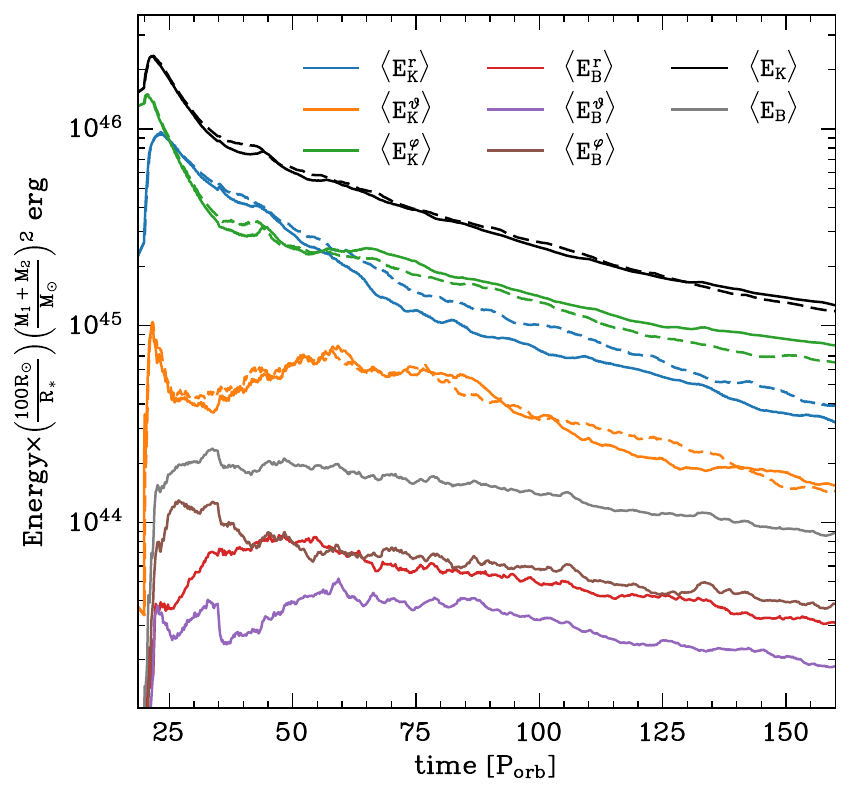} \hfill
         \caption{Evolution of the total kinetic and magnetic energy and of their radial, latitudinal and azimuthal components for \cite{Gagnier2023}'s simulation A (dashed lines) and this work's simulation B (solid lines).}
 \label{fig:Energy}
\end{figure}

We show the evolution of the magnetic energy
\begin{equation}\label{eq:Eb}
   \langle E_B \rangle = \int \frac{1}{2} \bB \cdot \bB \dd V
\end{equation}
and kinetic energy
\begin{equation}\label{eq:Eb1}
   \langle E_K \rangle = \int \frac{1}{2} \rho \buu\cdot \buu \dd V \ ,
\end{equation}
for simulation B in Fig.~\ref{fig:Energy}. We find that, after a rapid amplification phase, the total magnetic energy saturates and then slowly decays. The initial rapid amplification of the magnetic energy results from the stretching, folding, and winding of the initial weak poloidal field by differential rotation and turbulence. 

\begin{figure} 
\centering
        \includegraphics[ width=88mm]{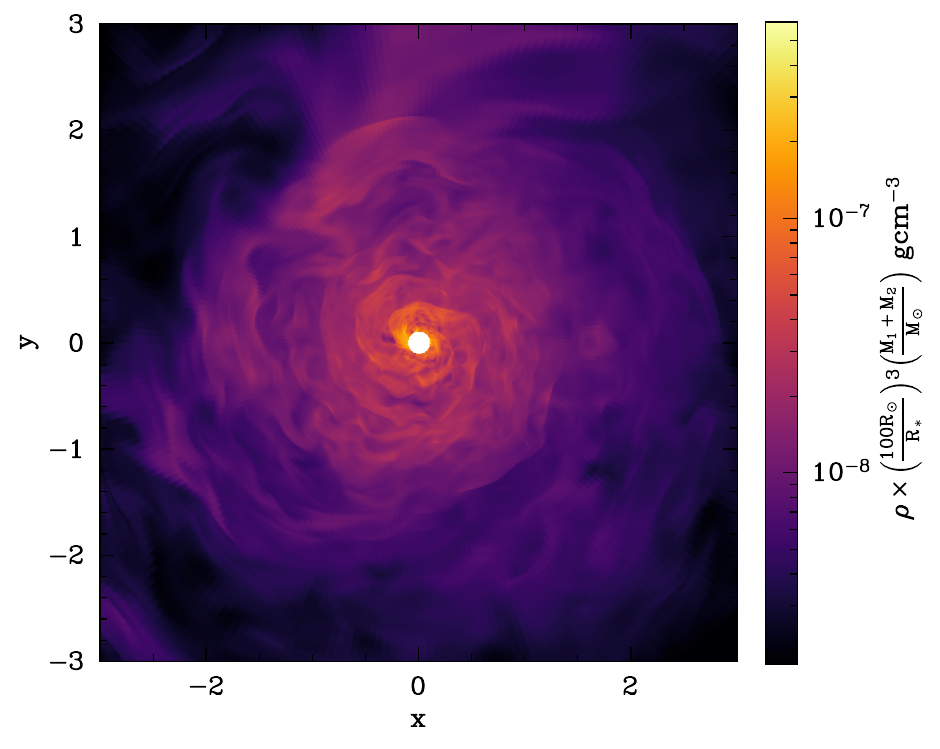} \\
            \includegraphics[ width=88mm]{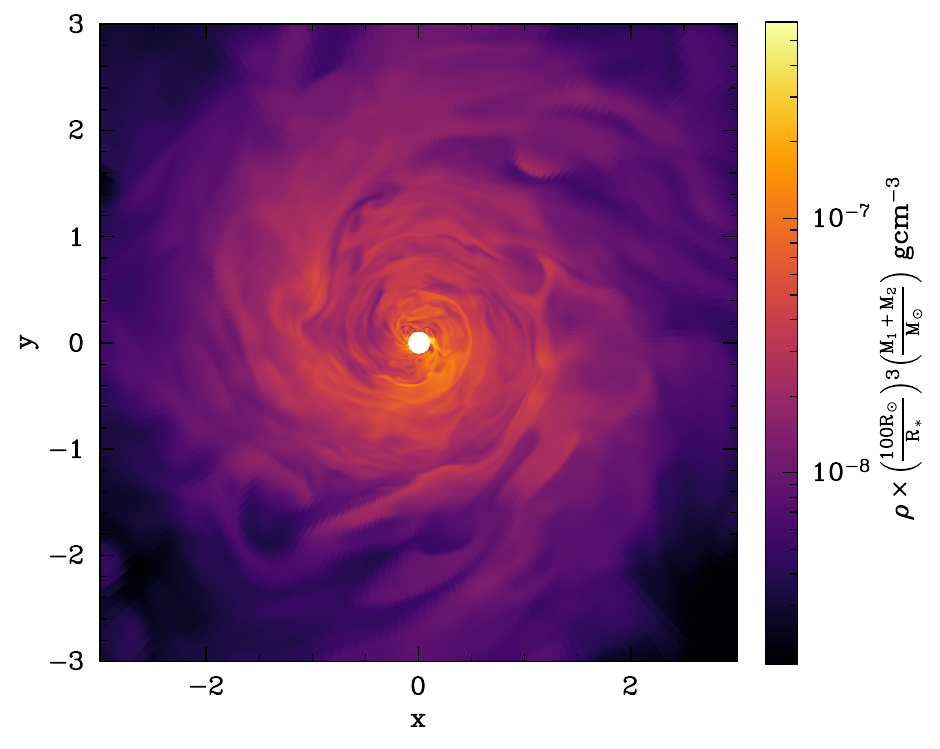} \hfill
   \caption{Zoomed-in snapshot of density cross section in the $xy$ plane at $t=140P_\text{orb}$ for \cite{Gagnier2023}'s simulation A (top) and this work's simulation B (bottom).}
\label{fig:snap}
\end{figure}

As the magnetic field strengthens, the Lorentz force begins to act back on the fluid, promoting more ordered and stable flow and leading to magnetic energy saturation. The magnetic tension associated with the strong toroidal field resulting from the winding up of the initial poloidal field, maintains the spiral density waves' structure by resisting their deformation against radial perturbation. This can be seen in Fig.~\ref{fig:snap} where we compare density snapshots taken at $t=140P_\text{orb}$ for models A and B, revealing that model B exhibits sharper spiral waves and a more uniform density distribution compared to model A. Additionally, magnetic tension tends to align fluid motion with magnetic field lines, resulting in a preference for azimuthal kinetic energy over radial kinetic energy during the saturated phase, in contrast to nonmagnetic simulations (Fig.~\ref{fig:Energy}). Hence, the presence of magnetic fields, even if they are relatively weak, has a significant impact on the envelope's structure and dynamics. This is discussed in more details later in this subsection, as well as in the following subsections, \ref{sec:KE} and \ref{sec:trans}. Finally, the magnetic energy at saturation, approximately $10^{44}$ erg, is similar to the findings from previous ab initio MHD simulations by \cite{Ohlmann2016b} and \cite{Ondratschek2022}. However, in our case, the ratio between kinetic and magnetic energy is 10, which is significantly smaller than in their results, where it was roughly 1000.

\begin{figure} 
\centering
      \includegraphics[ width=88mm]{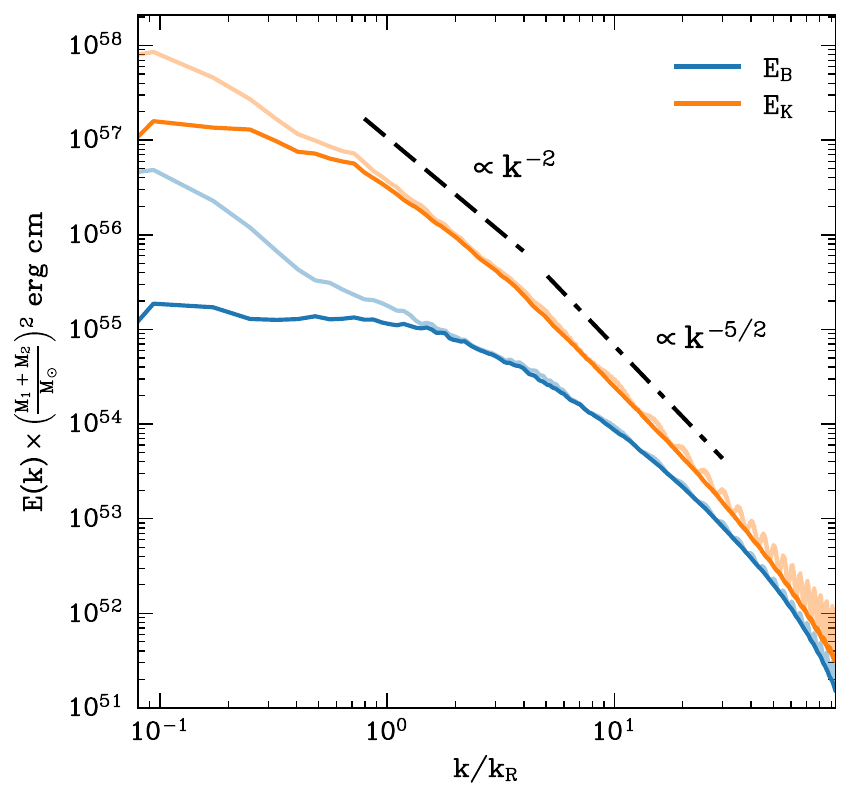} \hfill
   \caption{Kinetic and magnetic horizontally integrated energy spectra during the saturation phase obtained with $\ell_\text{max} = 128$ for simulation run B. Opaque lines represent the spectra excluding $\ell = 0$ components, while transparent lines include them. The oscillations visible at high $k/k_R$ when including $\ell = 0$ components result from the our direct integration of Eq.~(\ref{eq:Hankel}) which involves spherical Bessel functions $j_\ell(kr)$ that are very oscillatory for $kr \gg \ell$.}
\label{fig:Energy_spec}
\end{figure}

In Fig.~\ref{fig:Energy_spec}, we show the kinetic and magnetic horizontally integrated energy spectra averaged over twenty spectra spanning three orbital periods between $t= 137P_{\rm orb}$ and $t= 140P_{\rm orb}$ during the saturated phase \citep[see Appendix~\ref{app:ES} and e.g.,][]{Baddour2010,Parkin2013}. We find that kinetic energy is dominant on all scales with a diminishing ratio towards smallest scales where equipartition is approached.
To assess the relevant physics associated with the amplification and saturation of the magnetic energy, we write the total magnetic energy evolution equation (see Appendix~\ref{app:EB}), 
\begin{equation}\label{eq:Eb_ev}
\begin{aligned}
\langle     \dot{E}_B \rangle &=   \int \left(\bB \otimes \bB \right) {:} \bnabla \buu \dd V - \int E_B \bnabla \cdot \buu \dd V - \oint E_B \buu \cdot \boldsymbol{n} \dd S  \ , \\ 
    & = \langle \dot{E}_{B, \text{stretch}}\rangle + \langle \dot{E}_{B, \text{exp}} \rangle + \langle \dot{E}_{B, \text{adv}} \rangle \ ,
\end{aligned}   
\end{equation}
where $\boldsymbol{n}$ is the outward-pointing unit vector at the domain boundaries surface, $\langle \dot{E}_{B, \text{stretch}}\rangle$ is associated with the change in magnetic energy within the domain resulting from the stretching of the magnetic field lines by the fluid elements, $ \langle \dot{E}_{B, \text{exp}} \rangle$ corresponds to changes in magnetic energy due to expansion or compression effects, and $ \langle \dot{E}_{B, \text{adv}} \rangle$ corresponds to changes in magnetic energy from field advection through the domain boundaries. 

\begin{figure} 
 \centering
       \includegraphics[ width=88mm]{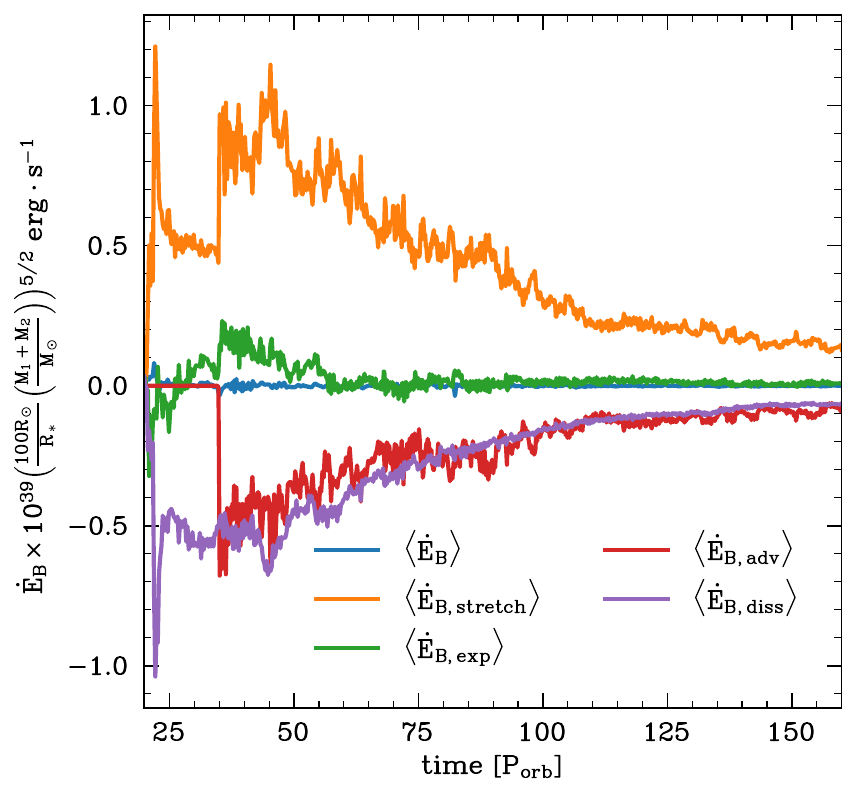} \hfill
    \caption{Evolution of the rate of change of magnetic energy $\langle \dot{E}_B \rangle$ and of its components defined in Eq.~(\ref{eq:Eb_ev}), for simulation run B. The numerical dissipation rate of magnetic energy by Joule heating is measured as $\langle \dot{E}_{B, \rm diss} \rangle$ = $\langle \dot{E}_{B} \rangle$ - $\langle \dot{E}_{B, \rm stretch} \rangle$ - $\langle \dot{E}_{B, \rm exp} \rangle$ - $\langle \dot{E}_{B, \rm adv} \rangle$.}
 \label{fig:dEmag}
 \end{figure}

\begin{figure} 
\centering
      \includegraphics[ width=88mm]{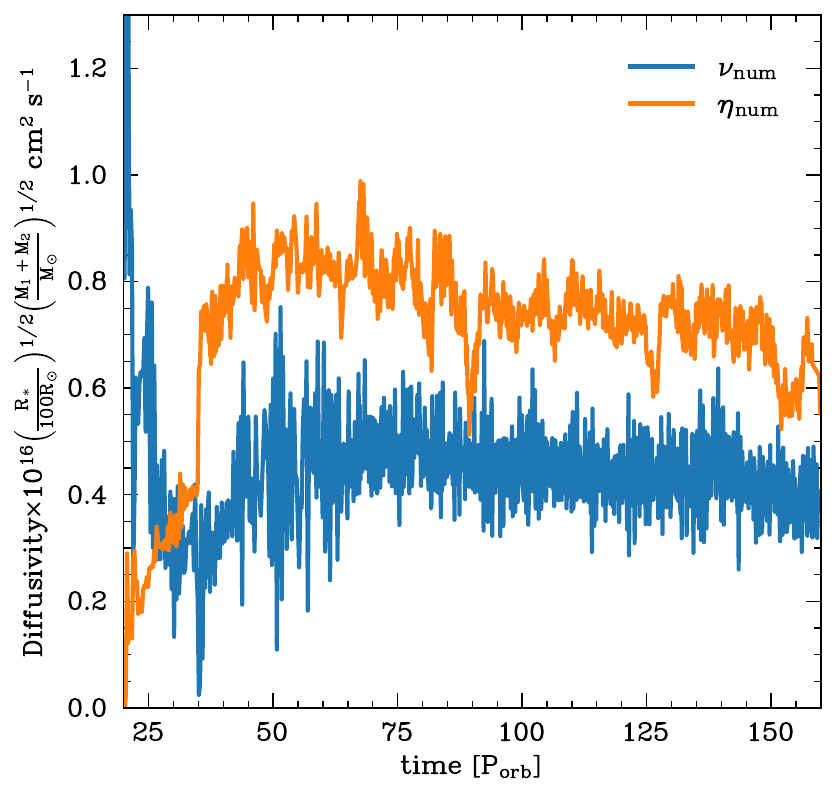} \hfill
   \caption{Time evolution of the measured numerical resistivity (Eq.~(\ref{eq:etanum})) and viscosity (Eq.~(\ref{eq:nunum})) for simulation run B. At quasi-steady state, we find $\nu_{\rm num} \simeq 4.35\times 10^{15}\ \rm  cm^2 s^{-1}$ and  $\eta_{\rm num} \simeq 7.11\times 10^{15}\ \rm cm^2 s^{-1}$.  }
\label{fig:nueta}
\end{figure}

The difference between the direct measurement of  $\langle \dot{E}_{B} \rangle$ and of the sum $ \langle \dot{E}_{B, \text{stretch}} \rangle + \langle \dot{E}_{B, \text{exp}} \rangle + \langle \dot{E}_{B, \text{adv}} \rangle$ corresponds to the numerical dissipation rate of magnetic energy by Joule heating $\langle \dot{E}_{B,\text{diss}} \rangle$. Assuming numerical dissipation to be of the Ohmic form \citep[e.g.,][]{Parkin2013}, we estimate the numerical resistivity as
\begin{equation}\label{eq:etanum}
    \eta_\text{num} = \frac{\langle \dot{E}_{B,\text{diss}} \rangle}{\langle \bB \Delta\bB \rangle} = -\frac{ \langle \dot{E}_{B,\text{diss}} \rangle}{\oint \left( \bnabla \cdot \bsigma \right) \cdot \boldsymbol{n} \dd S + \int |\bnabla \times \boldsymbol{B}|^2 \dd V} \ .
\end{equation}
Here, $\bsigma = \bB \otimes \bB - E_B \boldsymbol{\rm I}$ is the total Maxwell stress tensor and we used integration by parts and the identity $\boldsymbol{B} \times \left( \bnabla \times \boldsymbol{B} \right) = \bnabla (\boldsymbol{B} \cdot \boldsymbol{B})/2 - (\boldsymbol{B} \cdot \bnabla) \boldsymbol{B}$. We show the different contributions to the evolution of the total magnetic energy density in Fig.~\ref{fig:dEmag} and our measure of the numerical resistivity $\eta_{\rm num}$ in Fig.~\ref{fig:nueta}. We find a magnetic Reynolds number $\text{Re}_M^{\text{QSS}} = \mathcal{O}(10^3)$ at quasi-steady state for our models (see Table~\ref{tab:runs}). However, we emphasize that the Ohmic dissipation formalism might not be  valid and the measured resistivity likely depends on grid resolution, time integration method, spatial reconstruction scheme and Riemann solver \citep[e.g.,][]{Rembiasz2017}. In Fig.~\ref{fig:dEmag}, we see that the dominant source of magnetic energy during both the growth and saturation phases is the stretching of the magnetic field lines by the velocity shear, which accounts for $95 \%$ of magnetic energy generation during QSS. The remaining $5\%$ is produced by compression against magnetic pressure. The stretching of the magnetic field lines and the convergent transport of magnetic energy transfer kinetic energy into magnetic energy and effectively maintains dynamo action against dissipation and advection through the boundaries.

\subsubsection{Kinetic energy budget}\label{sec:KE}
To assess the relevant physics associated with the evolution of the kinetic energy budget and the dynamical relevance of magnetic fields in our simulations, we decompose the  kinetic energy evolution equation into mean and turbulent contributions in Appendix~\ref{app:MTKE} \citep[see also][]{Favre1965,Favre1969}. The Reynolds averaged mean kinetic energy evolution equation reads
\begin{equation}\label{eq:Ekmean_rt}
\begin{aligned}
 &\overline{\dot{E}_K^\text{mean}(r,\theta)} =      \frac{1}{2}\frac{\partial \overline{\rho}(\but \cdot \but)}{\partial t}  =- \bnabla \cdot \left(\overline{\rho} \but \frac{(\but \cdot \but)}{2}\right) -   \bnabla \cdot \left(\overline{\rho} \widetilde{\bup\otimes \bup} \cdot \but \right) \\&+ \overline{\rho}  \widetilde{\bup\otimes \bup} {:} \bnabla \but - \but \cdot \bnabla \overline{P}  +  \but \cdot \left(\bnabla \cdot \overline{\bsigma}\right)-  \but\overline{ \cdot \rho \bnabla \Phi}+ \but \cdot \left(\bnabla \cdot \overline{\btau}\right) \ ,
\end{aligned}
\end{equation}
and the Reynolds averaged turbulent kinetic energy evolution equation reads
\begin{equation}\label{eq:Ekturb_rt}
\begin{aligned}
    &\overline{\dot{E}_K^\text{turb}(r,\theta)} = \frac{1}{2}\frac{\partial \overline{\rho}(\widetilde{\bup \cdot \bup})}{\partial t} = - \bnabla \cdot \left(\overline{\rho}  \but \frac{(\widetilde{\bup \cdot \bup})}{2}\right)
   + \bnabla \cdot \Biggl( \overline{\bsigma \cdot \bup}  \\&  -   \overline{\rho \bup \frac{(\bup \cdot \bup)}{2}} - \overline{P^\prime \bup} \Biggr)  - \overline{\rho}  \widetilde{\bup\otimes \bup} {:} \bnabla \but - \overline{\bsigma {:} \bnabla \bup} - \overline{\bup} \cdot \bnabla \overline{P} \\&+ \overline{P^\prime \bnabla \cdot \bup} - \overline{\rho \bup \cdot \bnabla \Phi^\prime} + \overline{\bup \cdot \left( \bnabla \cdot \btau \right)} \ ,
  \end{aligned}
\end{equation}
where $\btau=  \nu_{\text{num}} \rho \boldsymbol{\rm c}$ is the (numerical) viscous stress tensor, $ \boldsymbol{\rm c} = \bnabla \buu  + (\bnabla \buu)^\intercal - (2/3) (\bnabla \cdot \buu)\boldsymbol{\rm I} $ is  the shear tensor and $\nu_{\text{num}}$ is the numerical viscosity that we assume spatially constant. 
Finally, we integrate Eqs.~(\ref{eq:Ekmean_rt}) and (\ref{eq:Ekturb_rt}) over the meridional plane, and multiply the result by $2 \pi$ to obtain the mean and turbulent kinetic energy equations.
\begin{equation}\label{eq:Ekmeantot}
\begin{aligned}
 \langle \dot{E}_K^\text{mean} \rangle  &= \langle \dot{E}_{K, \text{adv}}^\text{mean} \rangle + \langle \dot{E}_{K, \mathcal{R}}^\text{mean} \rangle + \langle \dot{E}_{K, \text{turb}}^\text{mean} \rangle + \langle \dot{E}_{K, P}^\text{mean} \rangle \\ & + \langle \dot{E}_{K, \mathcal{L}}^\text{mean} \rangle + \langle \dot{E}_{K, \Phi}^\text{mean} \rangle + \langle \dot{E}_{K, \text{diss}}^\text{mean} \rangle\ , \\ 
  \end{aligned}
 \end{equation}
where
\begin{subequations}
\begin{align}\label{eq:Ekmean_ev_1}
\langle \dot{E}_{K, \text{adv}}^\text{mean} \rangle &=  -     \Bigl\langle \bnabla \cdot \left(\overline{E_K^\text{mean}}  \but \right)\Bigr\rangle \ , \\
\langle \dot{E}_{K, \mathcal{R}}^\text{mean} \rangle &= -   \Bigl\langle    \bnabla \cdot \left(\overline{\rho} \but \cdot  \widetilde{\bup\otimes \bup}\right) \Bigr\rangle \ , \\
\langle \dot{E}_{K, \text{t}}^\text{mean} \rangle &=    \Bigl \langle  \overline{\rho} \widetilde{\bup\otimes \bup }{:} \bnabla \but  \Bigr \rangle \ , \\
\langle \dot{E}_{K, P}^\text{mean} \rangle &= -  \Bigl \langle\but \cdot \bnabla \overline{P} \Bigr \rangle \ , \\
\langle \dot{E}_{K, \mathcal{L}}^\text{mean} \rangle &=    \Bigl \langle \but \cdot \left(\bnabla \cdot \overline{\bsigma} \right) \Bigr\rangle \ , \\
 \langle \dot{E}_{K, \Phi}^\text{mean} \rangle &= -   \Bigl \langle  \but\cdot \overline{\rho \bnabla \Phi} \Bigr \rangle \ , \\ 
 \langle \dot{E}_{K, \text{diss}}^\text{mean} \rangle &=   \nu_{\text{num}} \Bigl\langle \but \cdot \left(\bnabla \cdot \overline{\rho \boldsymbol{\rm c}} \right) \Bigr\rangle \ .  \label{eq:Ekmean_ev_2}
\end{align}
\end{subequations}
Here, $\langle \dot{E}_{K, \text{adv}}^\text{mean} \rangle$ corresponds to the changes of mean kinetic energy from its advection through the domain boundaries, $\langle \dot{E}_{K, \mathcal{R}}^\text{mean} \rangle$ is associated with the transport of mean kinetic energy by Reynolds stress, $\langle \dot{E}_{K, \text{t}}^\text{mean} \rangle$ measures the destruction rate of mean kinetic energy into turbulence,  $\langle \dot{E}_{K, P}^\text{mean} \rangle$ is the work done by pressure on the mean flow, $\langle \dot{E}_{K, \mathcal{L}}^\text{mean} \rangle$ is the work done by the Lorentz force on the mean flow, and $\langle \dot{E}_{K, \Phi}^\text{mean} \rangle$ is the production rate of mean kinetic energy from the binary's gravitational potential energy. The turbulent kinetic energy evolution equation in turn reads
\begin{equation}\label{eq:Ekturbtot}
\begin{aligned}
 \langle \dot{E}_K^\text{turb} \rangle   &= \langle \dot{E}_{K, \text{adv}}^\text{turb} \rangle + \langle \dot{E}_{K, \mathcal{\text{trans}}}^\text{turb} \rangle + \langle \dot{E}_{K, \text{t}}^\text{turb} \rangle + \langle \dot{E}_{K, P}^\text{turb} \rangle \\ & + \langle \dot{E}_{K, \sigma}^\text{turb} \rangle + \langle \dot{E}_{K, \text{exp}}^\text{mean} \rangle \ ,
  \end{aligned}
 \end{equation}
 where
 \begin{subequations}
 \begin{align}\label{eq:Ekturb_ev_1}
\langle \dot{E}_{K, \text{adv}}^\text{turb} \rangle &=   -   \Bigl \langle \bnabla \cdot \left(\overline{E_K^\text{turb}}  \but \right) \Bigr \rangle \ ,\\
\langle \dot{E}_{K, \text{trans}}^\text{turb} \rangle &=    \Biggl \langle  \bnabla \cdot \Biggl( \overline{ \bsigma \cdot \bup}  -   \overline{\rho \bup \frac{(\bup \cdot \bup)}{2}} - \overline{P^\prime \bup }\Biggr) \Biggr \rangle \ , \\
\langle \dot{E}_{K, \text{t}}^\text{turb} \rangle &= -\langle \dot{E}_{K, \text{t}}^\text{mean} \rangle \ , \\
\langle \dot{E}_{K, P}^\text{turb} \rangle &= -   \Bigl \langle \overline{\bup} \cdot \bnabla \overline{P} \Bigr\rangle \ , \\
\langle \dot{E}_{K, \sigma}^\text{turb} \rangle &= -  \Bigl \langle \overline{\bsigma {:} \bnabla \bup } \Bigr \rangle \ , \\
\langle \dot{E}_{K, \text{exp}}^\text{turb} \rangle &= \Bigl \langle \overline{ P^\prime \bnabla \cdot \bup} \Bigr \rangle \ , \\
\langle \dot{E}_{K, \Phi}^\text{turb} \rangle &=   - \Bigl\langle \overline{\rho \bup \cdot \bnabla \Phi^\prime} \Bigr\rangle \ , \\
\langle \dot{E}_{K, \text{diss}}^\text{turb} \rangle &=   \nu_{\text{num}} \Bigl\langle \overline{ \bup \cdot \left(\bnabla \cdot \rho \boldsymbol{\rm c} \right)}\Bigr\rangle \ . \label{eq:Ekturb_ev_2}
\end{align}
 \end{subequations}
Here, $\langle \dot{E}_{K, \text{adv}}^\text{turb} \rangle$  corresponds to the changes in turbulent kinetic energy from its advection through the domain boundaries, $\langle \dot{E}_{K, \text{trans}}^\text{turb} \rangle$ is the turbulent kinetic energy transport rate from the work done by the Lorentz and pressure forces and by turbulent velocity fluctuations, $\langle \dot{E}_{K, \text{t}}^\text{turb} \rangle$  measures the production of turbulent kinetic energy from the interaction between mean shear and Reynolds stress, $\langle \dot{E}_{K, P}^\text{turb} \rangle$  is the work done by pressure on the turbulent flow, $\langle \dot{E}_{K, \sigma}^\text{turb} \rangle$ is the rate of production or destruction of turbulent kinetic energy from the interaction between turbulent shear and Maxwell stress, and $\langle \dot{E}_{K, \text{exp}}^\text{mean} \rangle$ is the rate of change of turbulent kinetic energy due to pressure dilatation. Finally $\langle \dot{E}_{K, \Phi}^\text{turb} \rangle$  is the production rate of turbulent kinetic energy from the work done by the multipole moments of the binary's gravitational force on the turbulent flow. We note that because we apply Favre decomposition to the velocity field, the rate of change of the total kinetic energy is simply the sum of its mean and turbulent counterparts, that is, $\langle \dot{E}_K \rangle = \langle \dot{E}_K^\text{mean} \rangle + \langle \dot{E}_K^\text{turb} \rangle$. 

\begin{figure} 
\centering
      \includegraphics[ width=88mm]{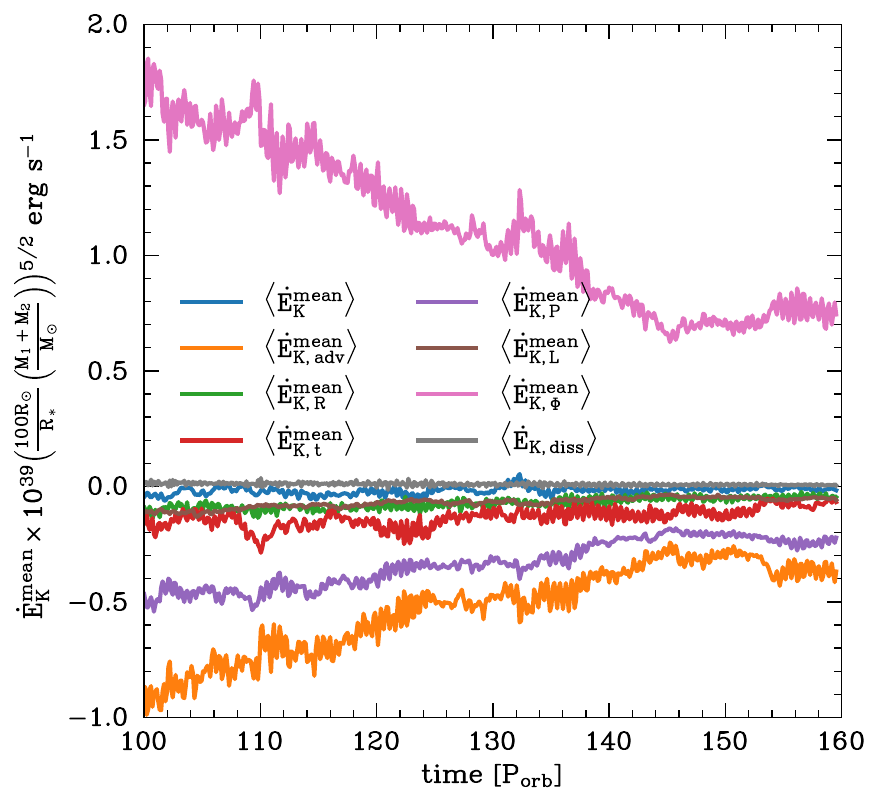} \hfill
   \caption{Evolution of the rate of change of mean kinetic energy $\langle \dot{E}_K^\text{mean} \rangle$ and of its components defined in Eqs.~(\ref{eq:Ekmean_ev_1})--(\ref{eq:Ekmean_ev_2}), for simulation run B.}
\label{fig:dEkmean_qss}
\end{figure}

\begin{figure} 
\centering
      \includegraphics[ width=88mm]{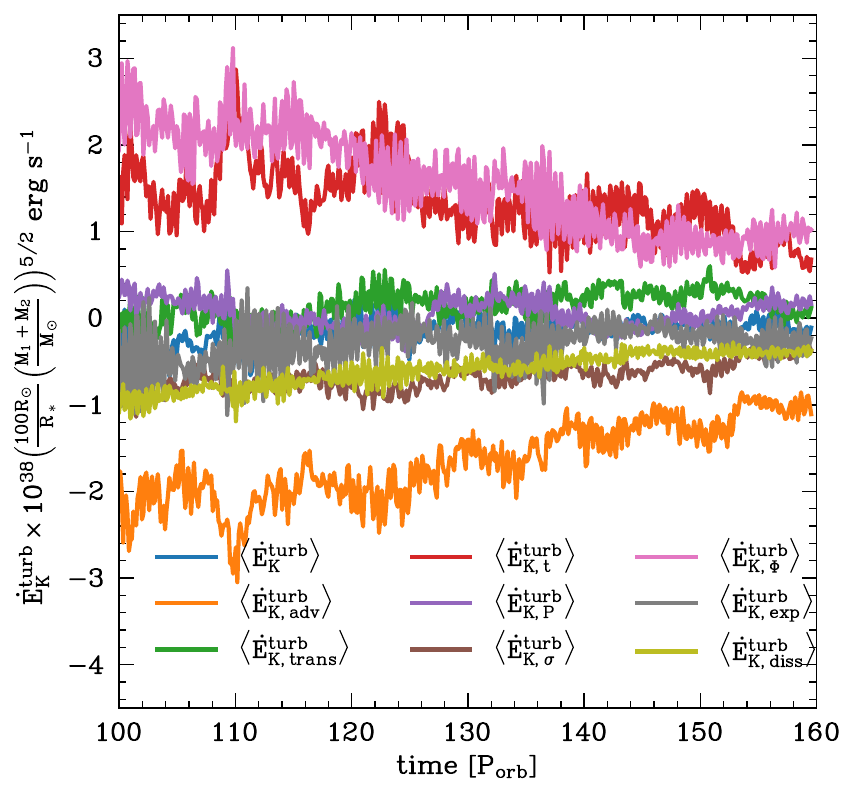} \hfill
   \caption{Evolution of the rate of change of turbulent kinetic energy $\langle \dot{E}_K^\text{mean} \rangle$ and of its components defined in Eqs.~(\ref{eq:Ekturb_ev_1})--(\ref{eq:Ekturb_ev_2}), for simulation run B.}
\label{fig:dEkturb_qss}
\end{figure}

We show the evolution of the  mean and turbulent kinetic energy change rates and of their components at QSS in Figs.~\ref{fig:dEkmean_qss} and~\ref{fig:dEkturb_qss}.
At QSS, we find that the binary orbital energy is the sole contributor to the envelope's mean kinetic energy production, while other processes act as sinks of mean kinetic energy. The primary sinks for mean kinetic energy are the losses through the boundaries and the work done by pressure on the mean flow. Additionally, but to a lesser extent, mean kinetic energy is dissipated into turbulence and lost due to the advection of mean flow by the Reynolds stress and by the work done by the Lorentz force.
Turbulent kinetic energy production is evenly split between the work done by the multipole moments of the binary's gravitational force on the turbulent flow and the interaction between the mean shear and Reynolds stress. Conversely, the loss of turbulent kinetic energy mainly comes from its advection through the domain boundaries. Still, numerical dissipation, pressure dilatation, and interactions between turbulent shear and Maxwell stress significantly contribute to the loss of turbulent kinetic energy.

Similarly to the way we measured numerical Ohmic resistivity, we estimate the numerical kinematic viscosity, assuming it to be spatially constant, as
\begin{equation}\label{eq:nunum}
\nu_\text{num} =\frac{\langle \dot{E}_{K, \text{diss}}^\text{mean} \rangle + \langle \dot{E}_{K, \text{diss}}^\text{turb} \rangle}{ \Bigl\langle \but \cdot \left(\bnabla \cdot \overline{\rho \boldsymbol{\rm c}} \right) \Bigr\rangle +\Bigl\langle \overline{ \bup \cdot \left(\bnabla \cdot \rho \boldsymbol{\rm c} \right)}\Bigr\rangle  }\ , 
\end{equation}
where $\langle \dot{E}_{K, \text{diss}}^\text{mean} \rangle$ and $\langle \dot{E}_{K, \text{diss}}^\text{turb} \rangle$ are obtained by substracting the various nondissipative components to the measured $\langle \dot{E}_{K}^\text{mean} \rangle$ and $\langle \dot{E}_{K}^\text{turb} \rangle$. We show our measurement of the numerical kinematic viscosity for simulation B in Fig.~\ref{fig:nueta}. We estimate the magnetic Prandtl number to be $P_M \simeq 0.6$, a value identical to that obtained by \cite{Parkin2014} in the context of global simulations of magnetorotational turbulence using the PLUTO code \citep{PLUTO2012}. However, astrophysical bodies are characterized by a wide range of magnetic Prandtl numbers which often substantially deviate from unity\citep[e.g.,][]{brandenburg2005,Balbus2008,brandenburg2009}. Our measured $P_M$ may thus not be realistic for common envelopes and cast doubts on the actual saturation values of $\alpha_M$ and $\alpha_P$, that is, on angular momentum transport efficiency. Still, an asymptotic regime $R_M > R_c$ in which $\alpha_P$ becomes $P_M$--independent may be reached in our simulations \citep[e.g.,][]{brandenburg2009,Kapyla2011a,Oishi2011}. Here, $R_M = VL/\eta_\text{num}$ is the magnetic Reynolds number quantifying the relative importance of magnetic induction to magnetic diffusion, $R_c$ is a critical Reynolds number of order $10^3$, and $V$ and $L$ are respectively a typical velocity and lengthscale of the flow. We reiterate here that our measure of the Prandtl number is only indicative and should be taken with caution. It is the result of strong approximations on the spatial uniformity of the numerical diffusivities and that the numerical diffusivity is equivalent to an Ohmic resistivity.

\subsubsection{Energy transfer analysis}\label{sec:trans}
Here we analyze the scales at which the energy sources described in sections~\ref{sec:ME} and \ref{sec:KE} operate and we focus attention to the scales at which magnetic energy is amplified. In order to do so, we perform a spectral energy transfer analysis, which is a commonly used method for understanding turbulent processes \citep[e.g.,][]{kraichnan67,Debliquy2005,Alexakis2007,Simon2009,PG2010,Lesur2011,Rempel2014,Grete2017,Grete2021,Du2023}. We show the energy transfer equations and the resulting transfer functions in Appendix \ref{app:TF}. We emphasize that our approach involves horizontally integrated Fourier transformation (see Appendix~\ref{app:ES}) for computing the transfer functions. Consequently, we are unable to assess the anisotropy of energy transfer in this analysis.

\begin{figure} 
      \includegraphics[width=88mm]{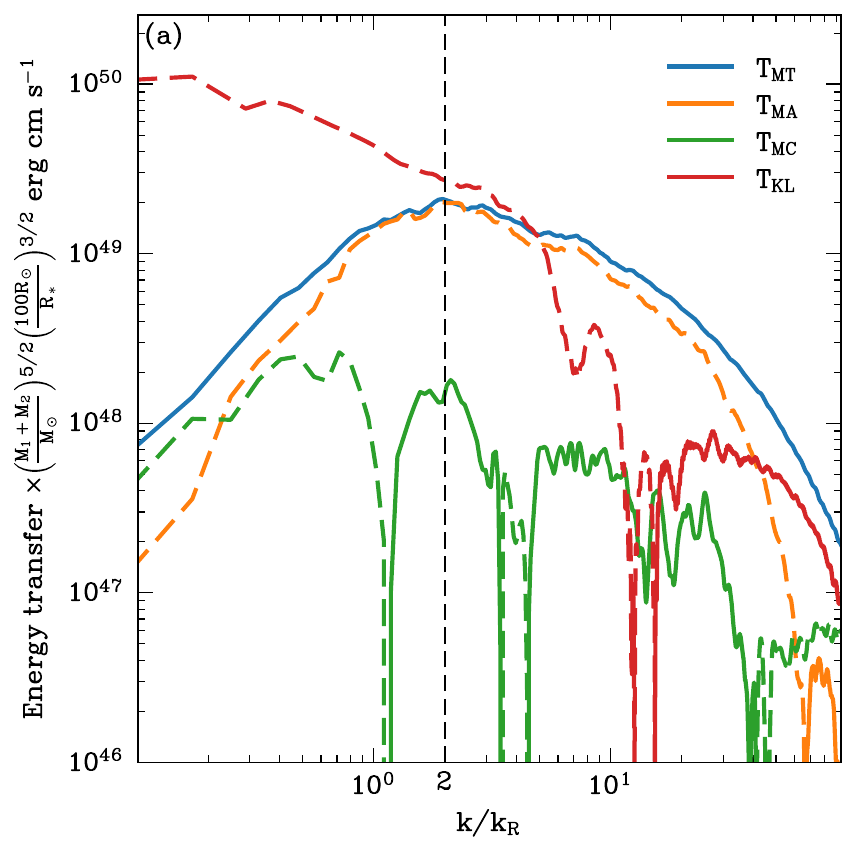} \\
      \includegraphics[width=88mm]{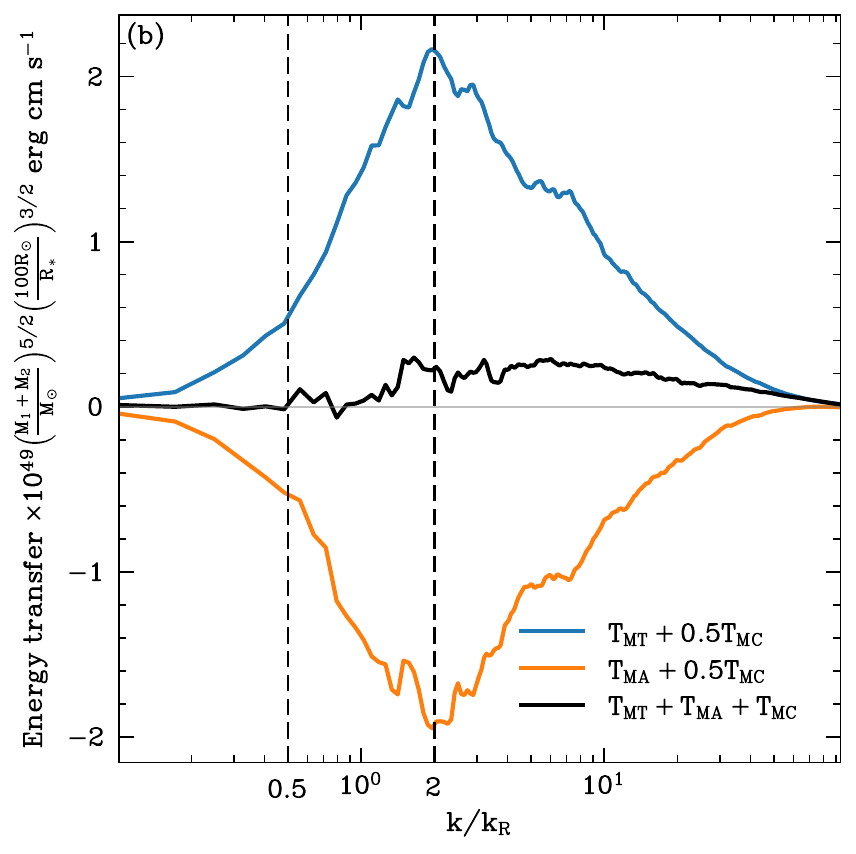} \hfill
   \caption{Results of the energy transfer analysis. Panel (a): Magnetic energy transfer as a function of the normalized wavenumber with $k_R = 2\pi/R_\ast$ averaged over twenty spectra spanning three orbital periods during the saturated phase, $137 \le t/P_{\rm orb} \le 140$. Positive values of transfer functions at a given wavenumber $k$ (full line) indicate gain of energy at the scale $2\pi/k$, negative values (dashed line) indicate a loss of energy at this scale. Panel (b): Same as panel a, except splitting the magnetic energy transfer rate into magnetic cascade contribution $T_{\rm MA}(k) + 0.5T_{\rm MC}(k)$ and kinetic to magnetic energy transfer $T_{\rm MT}(k) + 0.5T_{\rm MC}(k)$. 
}
\label{fig:transfer}
\end{figure}

In Fig.~\ref{fig:transfer}, we show the energy transfer quantities averaged over $137 \le t/P_{\rm orb} \le 140$. In panel (a), we display the magnetic energy transfer spectra associated with the stretching of magnetic field lines $T_\text{MT}(k)$, the advection of magnetic energy within the magnetic energy reservoir $T_\text{MA}(k)$, and with compression effects  $T_\text{MC}(k)$. In panel (b), we show the spectra associated with the magnetic cascade $T_{\rm MA}(k) + 0.5T_{\rm MC}(k)$ and the kinetic to magnetic energy transfer  $T_{\rm MT}(k) + 0.5T_{\rm MC}(k)$. Positive values of transfer functions at a given wavenumber $k$ indicate gain of energy at a scale $2\pi/k$. Conversely, negative values indicate a loss of energy at this scale. 
In panel (a), we find that $T_{\rm MT}(k)$ is positive for all wavenumbers $k$, indicating that magnetic energy production occurs through line stretching across all scales. Conversely, $T_{\rm MA}(k)$ is only positive on scales smaller than $\sim a_\text{b}/10$,  suggesting a forward magnetic energy cascade from larger to smaller scales through advection. Remarkably, both $T_{\rm MT}(k)$ and $T_{\rm MA}(k)$ peak at a scale of $R_\ast/2 \approx 3 a_\text{b}$, which is approximately the wavelength of the binary-driven spiral density waves, with almost equal but opposite values.
This indicates that the bulk of magnetic energy production occurs at that scale through line stretching and that the newly generated magnetic energy is promptly advected towards the smallest scales.
We find that the energy transfer rate by compression effects, $T_{\rm MC}(k)$, is overall weaker compared to $T_{\rm MT}(k)$ and $T_{\rm MA}(k)$. Unlike $T_{\rm MT}(k)$ and $T_{\rm MA}(k)$, transfer by compression effects $T_{\rm MC}(k)$ alternates between positive and negative values with varying $k$. This can be attributed to the compression--rarefaction pattern of spiral density waves. Although weaker, $T_{\rm MC}(k)$ plays a crucial role in the net magnetic energy production on the $R_\ast/2$ scale due to the near-perfect balance between $T_{\rm MT}(k)$ and $T_{\rm MA}(k)$ at this scale. In panel (b), we see that, on scales larger than $2 R_\ast$, $T_{\rm MT}(k) + 0.5T_{\rm MC}(k)$ is in balance with $T_{\rm MA}(k) + 0.5T_{\rm MC}(k)$. Such balance indicates no net gain of magnetic energy at scales larger than $2 R_\ast$ and thus implies the absence of large-scale magnetic field production. On smaller scales however, $T_{\rm MT}(k) + T_{\rm MA}(k) + T_{\rm MC}(k)$ is positive. Despite this, $\langle \dot{E}_B \rangle  \lesssim 0$ due to numerical dissipation of magnetic energy by Joule heating (see Figs.~\ref{fig:Energy} and \ref{fig:dEmag}). 

In panel (a) of Fig.~\ref{fig:transfer}, we also show the time-averaged spectra associated with the gain and loss of kinetic energy from the work of the Lorentz force via magnetic tension and magnetic pressure, $T_{\rm KL}$. The kinetic energy reservoir's perspective offers a different view. Indeed, we see that kinetic energy is lost to magnetic energy on scales down to $\sim 0.4 a_\text{b}$.  Approximately $20\%$ of the energy drawn from the kinetic reservoir on these larger scales returns as kinetic energy on smaller scales. Remarkably, we find that most of the kinetic energy is lost to magnetic energy on the largest scales. However, the gain of magnetic energy on these scales is notably lower, which hinders the generation of significant large-scale magnetic fields during QSS. This highlights the nonlocality of interactions between magnetic and kinetic energy in spectral space \citep{Alexakis2007}. Unfortunately, we cannot pinpoint the specific origin or destination scales in spectral space. This would require a more complex shell-to-shell energy transfer analysis \citep[e.g.,][]{Alexakis2007,Lesur2011,Grete2017,Grete2021} that is beyond the scope of this work. 

The lack of magnetic energy production on large radial scales implies that the amplified magnetic field may not be able to funnel and collimate a radial polar outflow, potentially impeding the formation of well-defined bipolar jets or jet-like outflows that often give rise to the characteristic bipolar shapes commonly observed in planetary nebulae \citep[e.g.,][]{GarciaSegura1999,GarciaSegura2018,GarciaSegura2020,Zou2020,Ondratschek2022}.
Nonetheless, the presence of a strong toroidal magnetic field about the orbital plane introduces magnetic tension. This tension acts to decelerate the expansion of the envelope in directions perpendicular to the polar axis. Additionally, centrifugal forces, along with turbulent mixing, jointly contribute to the formation of intermittent, jittering low-density regions near the poles (see Fig.~\ref{fig:snap925}) which may facilitate the efficient channeling of outflows. Hence, the combined effects of magnetic tension, centrifugal forces, and turbulent mixing may ultimately contribute to the formation of nonspherical planetary nebulae.

\subsection{Binary evolution and angular momentum conservation}
\label{sec:binary_evolution}

\begin{figure}
\centering
      \includegraphics[ width=88mm]{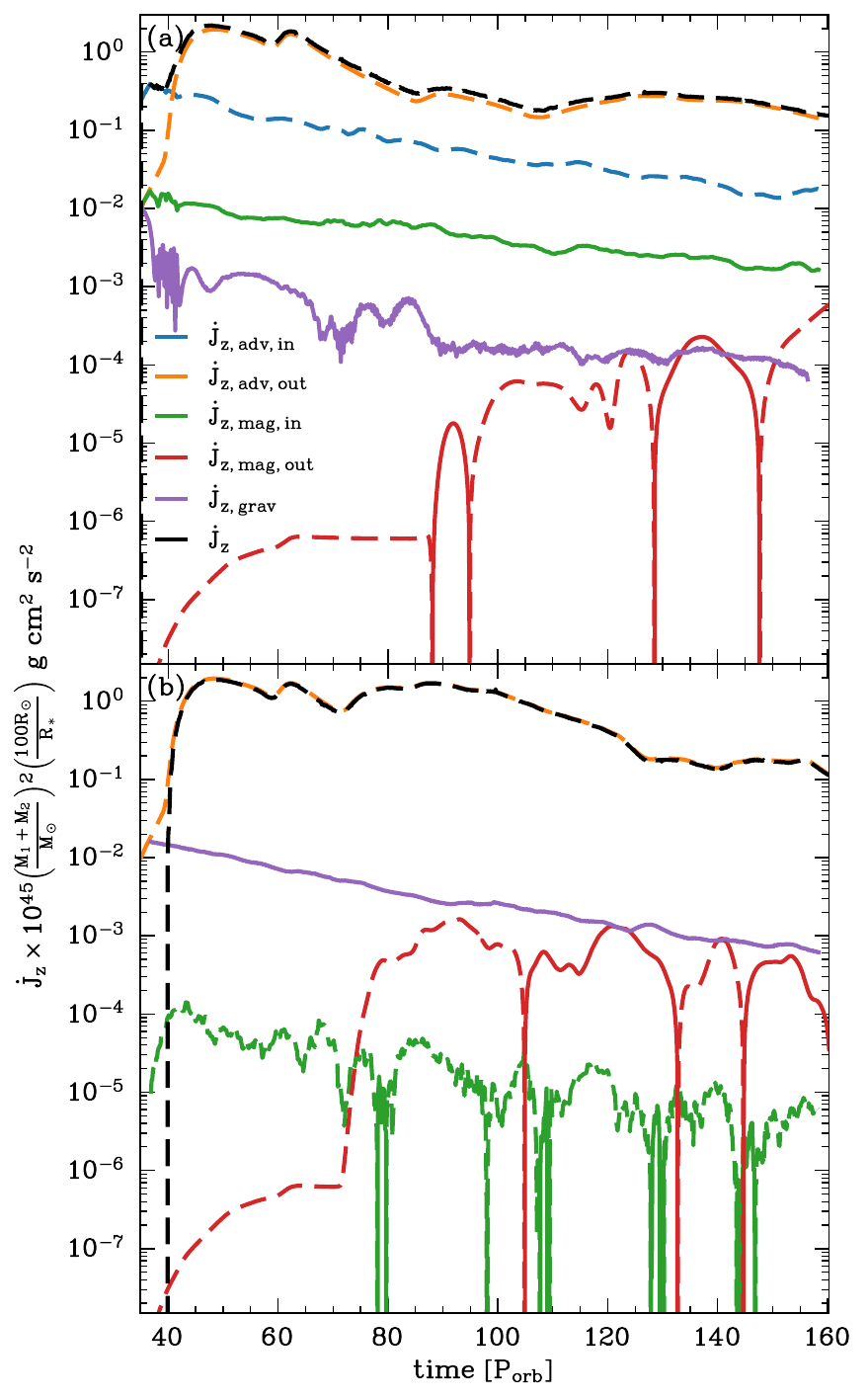} \hfill
   \caption{Moving average of the time evolution of the advective, magnetic, and gravitational torques for run B (panel a) and B' (panel b). The measured total torque is indicated by a black line, positive values are denoted by solid lines and negative values by dashed lines.}
\label{fig:Jz}
\end{figure}

Here, we address the impact of magnetic fields on the orbital evolution of the central binary. The commonly accepted model of common envelope evolution assumes a near-monotonic decrease in binary separation over time.  In \cite{Gagnier2023}, we showed that this process is characterized by an orbital contraction timescale that can be much shorter than the thermal timescale of the envelope, particularly when the central binary is able to accrete mass and angular momentum from the surrounding envelope. In our setup, and as in \cite{Gagnier2023}, we fix the orbital parameters and predict the orbital evolution by measuring the rate of angular momentum transfer between the binary and the envelope without self-consistently considering any reciprocal influence from the envelope to the binary orbit. To assess the influence of magnetic fields on the secular evolution of binary separation, it is crucial to evaluate the torques acting within the shared envelope. These torques originate from both the quadrupolar moment of the gravitational potential and the angular momentum fluxes resulting from stresses across the domain boundaries. The conservation equation for angular momentum can be expressed as
\begin{equation}
    \dot{J}_z = \dot{J}_{z, \rm adv} + \dot{J}_{z, \rm grav} + \dot{J}_{z, \rm mag} \ ,
\end{equation}
where $\dot{J}_{z, \rm adv}$ is the advective torque associated with the loss of angular momentum through the boundaries,
\begin{equation}
    \dot{J}_{z, \rm adv} = - \oint_{\partial R} \rho s u_\varphi\buu \cdot \boldsymbol{n}_\perp \dd S \ ,
\end{equation}
$\dot{J}_{z, \rm grav}$ is the gravitational torque exerted by the binary,
\begin{equation}
    \dot{J}_{z, \rm grav} =  -\int  \rho  s \frac{\partial \Phi}{\partial \varphi} \dd V \ ,
\end{equation}
and $\dot{J}_{z, \rm mag}$ is the magnetic torque,
\begin{equation}
    \dot{J}_{z, \rm mag} =  \oint_{\partial R}  s B_\varphi\boldsymbol{B} \cdot \boldsymbol{n}_\perp \dd S \ .
\end{equation}
Here, $\boldsymbol{n}_\perp$ is the outward-pointing unit vector at the boundaries' surface. In Fig.~\ref{fig:Jz}, we show the evolution of these torques for runs B and B'. We find that at QSS, as in the nonmagnetic simulations of \cite{Gagnier2023}, the total angular momentum evolution is primarily driven by the outflow at the outer boundary, which is a consequence of envelope expansion and the finite radial extent of our numerical domain. Mass accretion hinders material buildup at the inner boundary and facilitates efficient horizontal turbulent mixing \citep[as discussed in][]{Gagnier2023}. Such efficient mixing weakens the gravitational torque, resulting in a modest transfer rate of angular momentum from the binary's orbit to the envelope at QSS. Conversely, when accretion onto the binary is prevented (simulation run B'), material accumulation at the inner boundary and its stabilizing effect leads to an injection rate of angular momentum by the gravitational torque that is an order of magnitude larger than when accretion is allowed. The influence of magnetic fields on angular momentum evolution remains marginal in our two simulation runs. This is particularly evident in simulation run B' where magnetic fields cannot be advected through the inner boundary by the fluid flow, unlike in simulation run B. Nevertheless, this impact very much depends on our selection of boundary conditions for the magnetic field.

\begin{figure} 
\centering
      \includegraphics[ width=88mm]{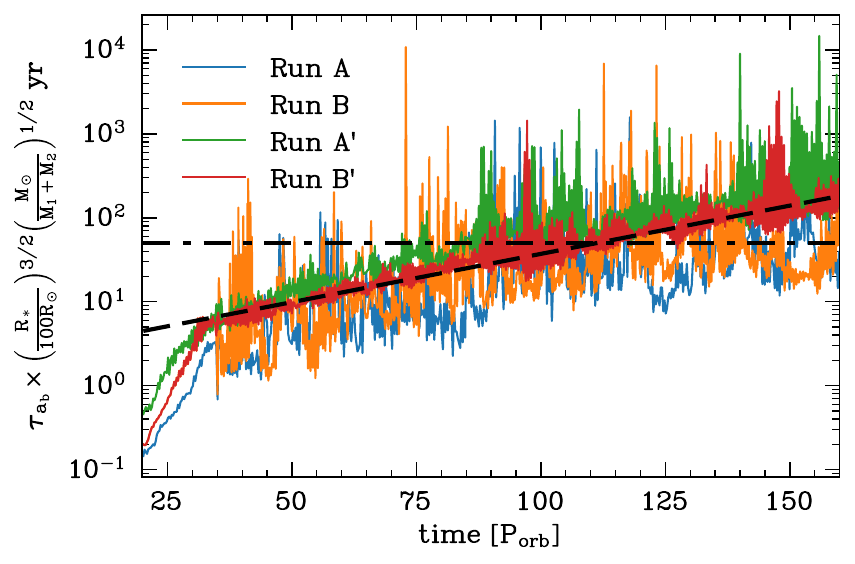} \hfill
   \caption{Time evolution of the orbital separation evolution timescale for our two MHD runs B and B', along with the non-MHD runs A and A' from \cite{Gagnier2023}. The black dashed line represents $\tau_{a_\text{b}} = 50\,\text{yr}$, while the black dashdotted line corresponds to $\tau_{a_{\rm b}} = \lambda 10^{\gamma t/P_\text{orb}}$ with $\lambda = 0.42$ and $\gamma = 0.0115$.}
\label{fig:tau_ab}
\end{figure}

As in \cite{Gagnier2023}, we set the orbital eccentricity $e_\text{b} = 0$ and enforce the binary mass ratio to $q = 1$. Consequently, we assume an equal distribution of mass and angular momentum accretion through the inner boundary between the two cores, resulting in $\dot{q} = 0$. Additionally, we make the assumption that $\dot{e}_\text{b} = 0$. The time derivative of the binary's angular momentum can then be expressed using the orbital separation evolution equation
 \begin{equation}\label{eq:adota}
 \frac{\dot{a_\text{b}}}{a_\text{b}} = \frac{\dot{M}}{M} \left( 2 \frac{M \dot{J}_{z,\text{b}}}{\dot{M}J_{z,\text{b}}} - 3\right) = \frac{1}{\tau_{a_\text{b}}} \ ,
 \end{equation}
 where $\tau_{a_\text{b}}$ is the orbital separation evolution timescale. If accretion onto the binary is disabled, $\dot{M} = 0$ and $\dot{J}_{z,\text{b}} = -\dot{J}_{z,\rm grav}$, otherwise $\dot{M} = -\int_{\partial R_{\rm in}} \rho u_r \dd S$ and $\dot{J}_{z,b} = -\left.\dot{J}_{z, \rm adv}\right\vert_{r = r_{\rm in}} - \dot{J}_{z, \rm grav} - \left.\dot{J}_{z, \rm mag}\right\vert_{r = r_{\rm in}}$. We show the orbital separation evolution timescale for our two MHD runs B and B’, along with the non-MHD runs A and A’ from \cite{Gagnier2023} in Fig.~\ref{fig:tau_ab}. As expected from the previous section, magnetic fields play little to no role in the binary separation evolution. Hence, as in the nonmagnetic simulations of \cite{Gagnier2023}, we find that the orbital separation evolution timescale reaches a quasi-steady value of the order $10$--$100$ years when accretion onto the binary is allowed, and increases with time as
 \begin{equation}
     \tau_{a_{\rm b}}\propto 10^{\gamma t/P_\text{orb}} \ ,
 \end{equation}
with $\gamma \simeq 0.0115$, when accretion onto the binary is prevented. 

\subsection{Angular momentum transport}\label{sec:AMtransport}

Radial transport of angular momentum within the common envelope can either facilitate or hinder the binary's post-dynamical inspiral through the generation of accretion flows. This process also shapes the envelope's structure by introducing density distribution asymmetries that enable angular momentum exchange between the binary and the envelope via the gravitational torque. These interactions further contribute to the shaping of emerging planetary nebulae resulting from the common envelope outflow. Magnetic fields are also well known to enhance or impede angular momentum transport within various astrophysical systems, as well as to generate torques, instabilities, and outflows, such as winds or jets. Here, we analyze angular momentum transport within the shared envelope, evaluating the morphology and strength of this process while assessing the influence of each physical mechanism.

\begin{figure} 
\centering
      \includegraphics[width=88mm]{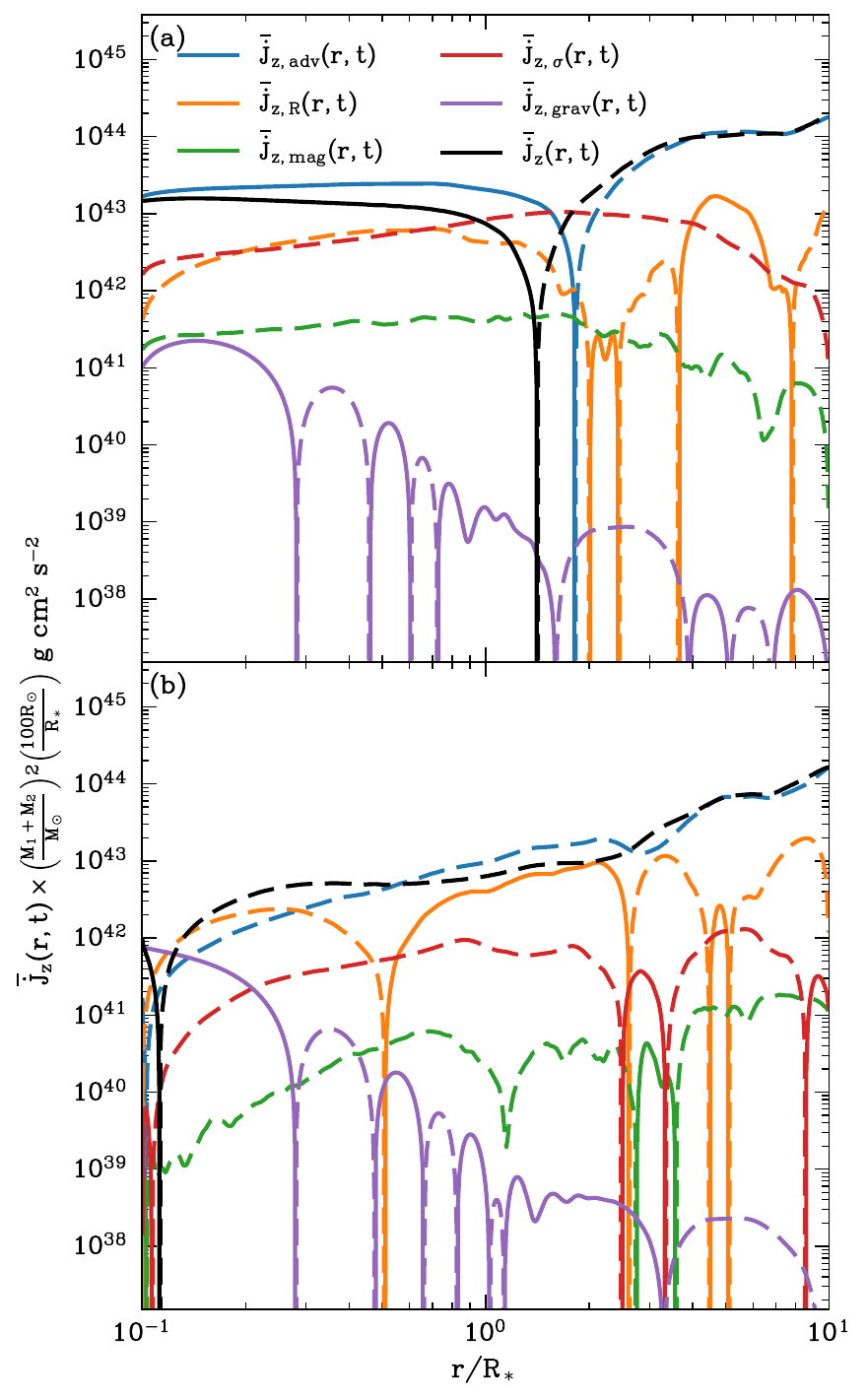} \hfill
   \caption{Gravitational torque and mean-flow and turbulent contributions to the local angular momentum transfer rate across the common envelope for model B (panel a) and B' (panel b), averaged in time from $t=140P_\text{orb}$ to $t=160P_\text{orb}$. Dashed lines depict negative values, whereas solid lines represent positive values.}
\label{fig:jdotprof}
\end{figure}

In Fig.~\ref{fig:jdotprof}, we show the radial profile of the gravitational torque and mean-flow and turbulent contributions to the local angular momentum transfer rate across the common envelope, averaged over $140 \le t/P_\text{orb} \le 160$ for simulations B (panel a) and B' (panel b). The transfer rates are defined as (see Eq.~(\ref{eq:AMcons}))
\begin{align}
    & \overline{\dot{J}}_{z, \rm adv}(r,t) = - \oint_{\partial r} s \overline{\rho} \,  \widetilde{u}_\varphi \widetilde{u}_r \dd S \ , \\
    & \overline{\dot{J}}_{z, \mathcal{R}}(r,t) = - \oint_{\partial r} s \overline{\rho}\,\widetilde{u_\varphi^{\prime\prime} u_r^{\prime\prime}} \dd S \ , \\    
    & \overline{\dot{J}}_{z, \rm mag}(r,t) =  \oint_{\partial r} s  \overline{B}_\varphi \overline{B}_r \dd S \ , \\
    & \overline{\dot{J}}_{z, \sigma}(r,t) =  \oint_{\partial r}  s  \overline{B_\varphi^{\prime} B_r^{\prime}} \dd S \ , \\        
    & \overline{\dot{J}}_{z, \rm grav}(r,t) = \int^{R_{\rm domain}}_r \left( \int_{\partial r} s \overline{\rho \frac{\partial \Phi^\prime}{\partial \varphi}} \dd S \right) \dd r \ . 
\end{align}

When accretion onto the binary is allowed (simulation run B), we find that angular momentum is transported inward for $r \lesssim 1.4R_\ast$, and outward further out in the envelope. It is dominated by the contribution from the mean-flow advection $\overline{\dot{J}}_{z, \rm adv}(r,t)$ which is directed inward for $r \lesssim 1.8R_\ast$, and outward further out. The transport of angular momentum associated with the mean magnetic field $\overline{\dot{J}}_{z, \rm mag}(r,t)$ is directed outward across the envelope, with negligible amplitude. Within $r \lesssim 0.75R_\ast$, turbulent Reynolds $\overline{\dot{J}}_{z, \mathcal{R}}(r,t)$ and Maxwell stresses $\overline{\dot{J}}_{z, \sigma}(r,t)$ contribute equally and significantly to outward angular momentum transport. Moving outward, the Reynolds stress changes direction intermittently, while the turbulent Maxwell stress maintains an outward direction with comparable magnitude.  Overall, we find the combined effect of turbulent Reynolds and Maxwell stresses on radial transport of angular momentum to be significant. The combined effect damps the inward transport of angular momentum driven by the mean flow by a factor $\sim 2$ for $r \lesssim 1.4R_\ast$, it reverses the direction of angular momentum transport for $ 1.4 \lesssim r/R_\ast \lesssim 1.8$, and the turbulent Maxwell stress slightly enhances outward transport for $ 1.8 \lesssim r/R_\ast \lesssim 4$.  Conversely, the  binary's gravitational torque impact on the radial angular momentum transport is negligible throughout the envelope. However, a notable feature is the presence of global radial oscillations about zero, which arise due to the gravitational coupling between the binary potential and the spiral density waves originating near the binary and propagating out in the envelope \citep{Cimerman2023}.

When accretion onto the binary is prevented (simulation run B'), angular momentum is transported outward across the envelope. Notably, the positive values of $\overline{\dot{J}}_z$ within the very inner envelope result from the injection of angular momentum by the binary's gravitational torque and they do not indicate inward transport. We further note that this injection of angular momentum is more pronounced in simulation B' compared to simulation B owing to material accumulation at the reflecting inner boundary.
In simulation B', and contrary to simulation B, magnetic fields have a negligible impact on radial angular momentum transport. The primary driver remains  advection by the mean flow, with the Reynolds stress also making a significant contribution. The Reynolds stress alternates between dampening and enhancing outward transport across varying radii from the central binary.

\begin{figure} 
\centering
      \includegraphics[ width=88mm]{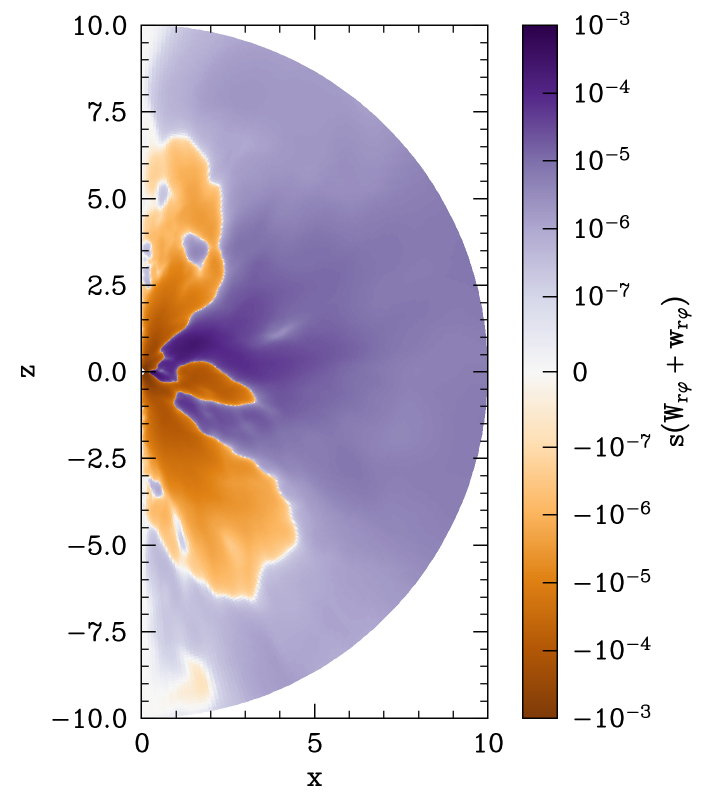} \hfill
   \caption{Reynolds average of the total radial angular momentum flux $s\mathcal{W}_{r\varphi}$, averaged over twenty orbital periods for $140 \leq t/P_{\rm orb} \leq 160$ for simulation B. }
\label{fig:AM_trans_full}
\end{figure}

\begin{figure} 
\centering
      \includegraphics[ width=88mm]{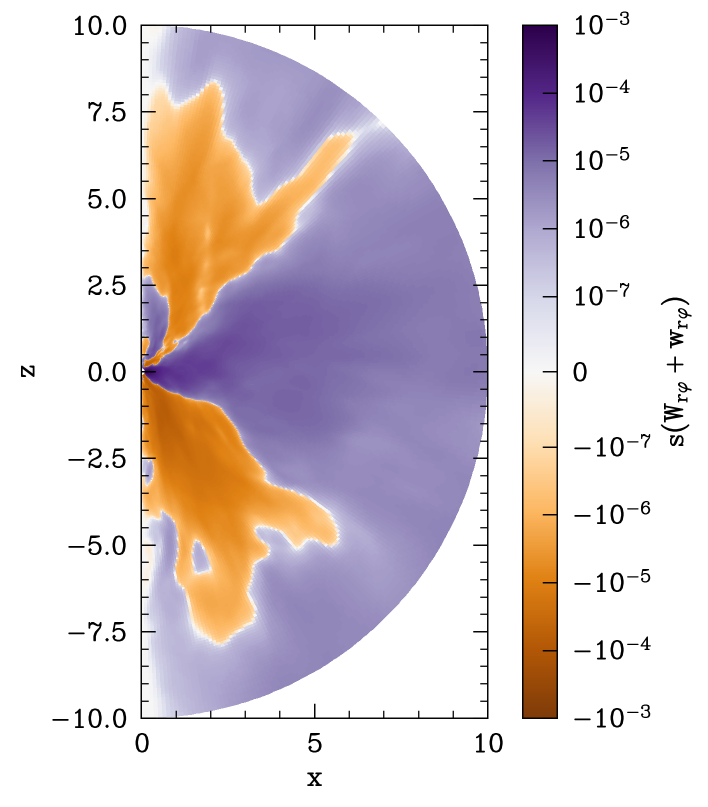} \hfill
   \caption{Same as Fig.~\ref{fig:AM_trans_full} but for simulation B'.}
\label{fig:AM_trans_full_ref}
\end{figure}

\begin{figure} 
\centering
      \includegraphics[ width=88mm]{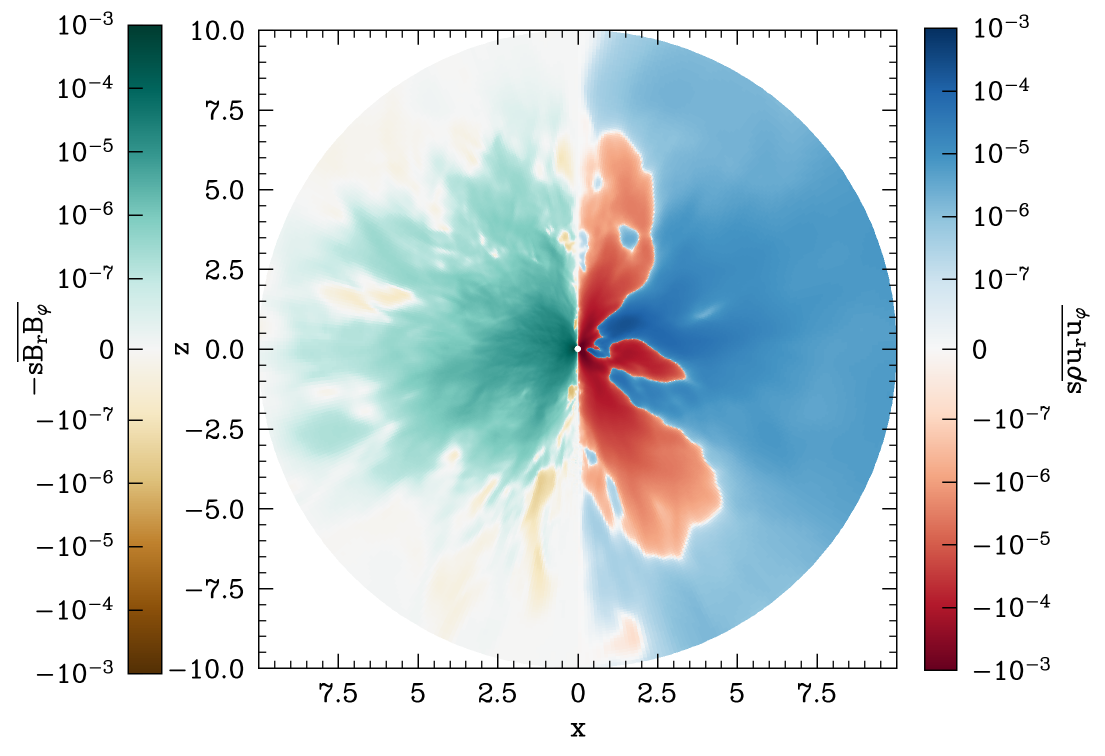} \\
          \includegraphics[ width=88mm]{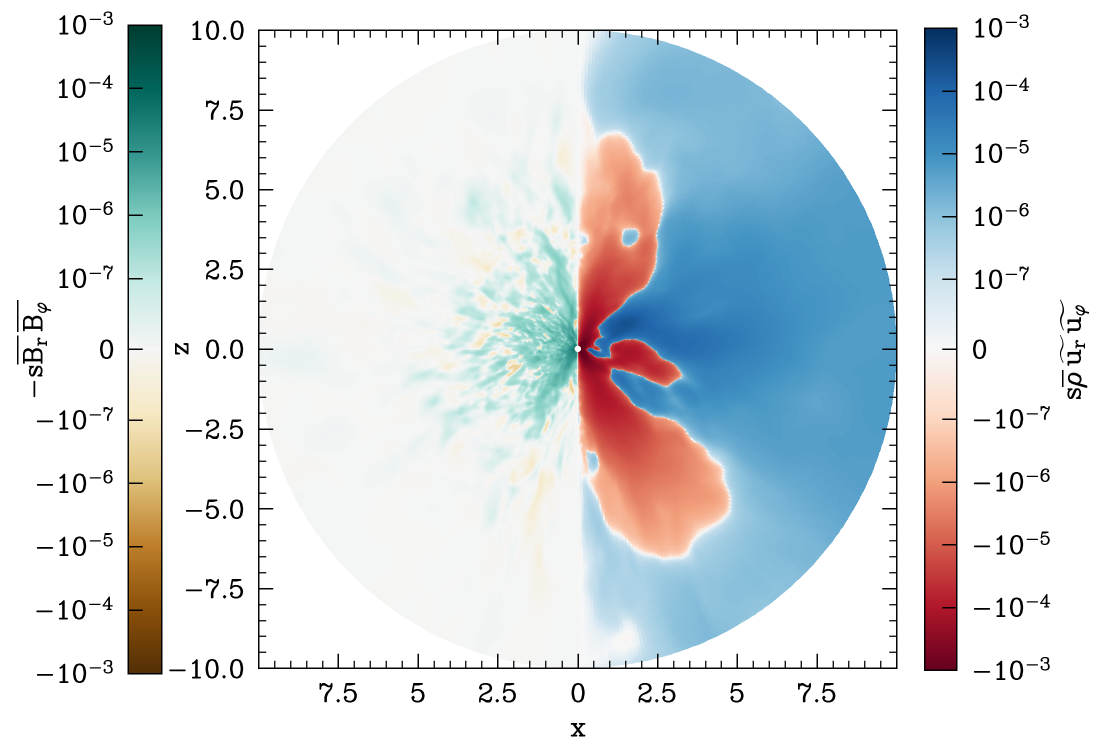} \\
              \includegraphics[ width=88mm]{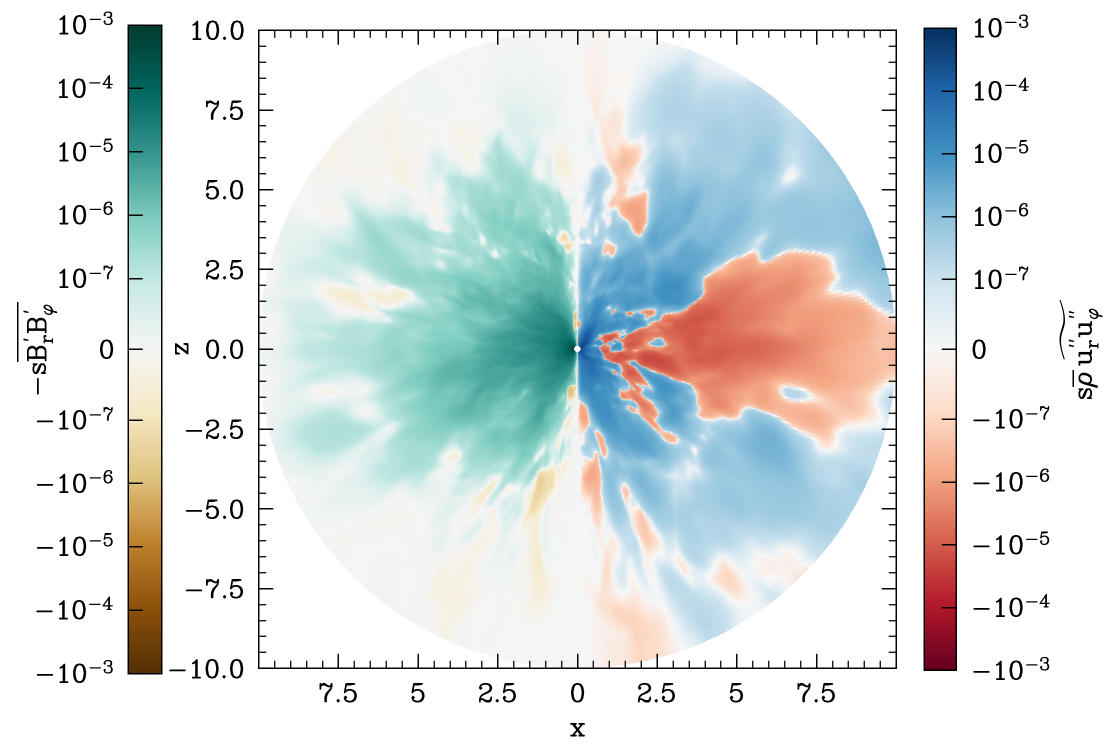} \hfill
   \caption{Reynolds average of the mean  and turbulent contribution of the of the total radial angular momentum flux (see Eqs.~(\ref{eq:comp1}) and (\ref{eq:comp2})), averaged over twenty orbital periods for $140 \leq t/P_{\rm orb} \leq 160$ for simulation B.}
\label{fig:AM_trans}
\end{figure}

\begin{figure} 
\centering
      \includegraphics[ width=88mm]{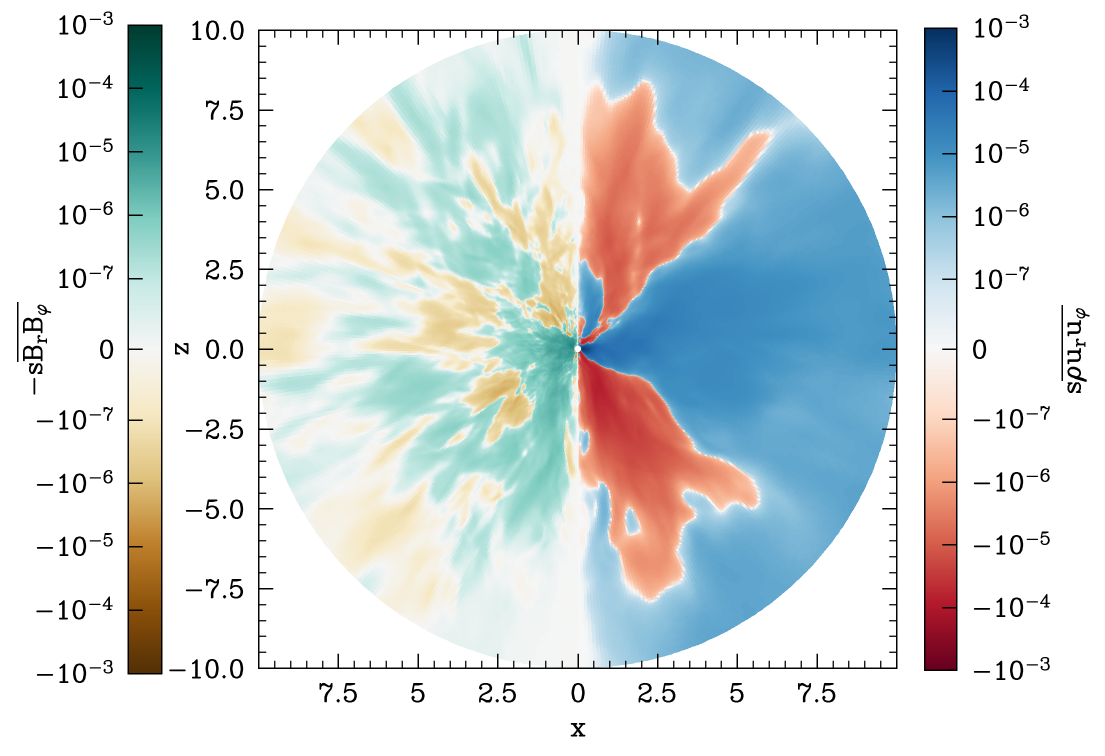} \\
          \includegraphics[ width=88mm]{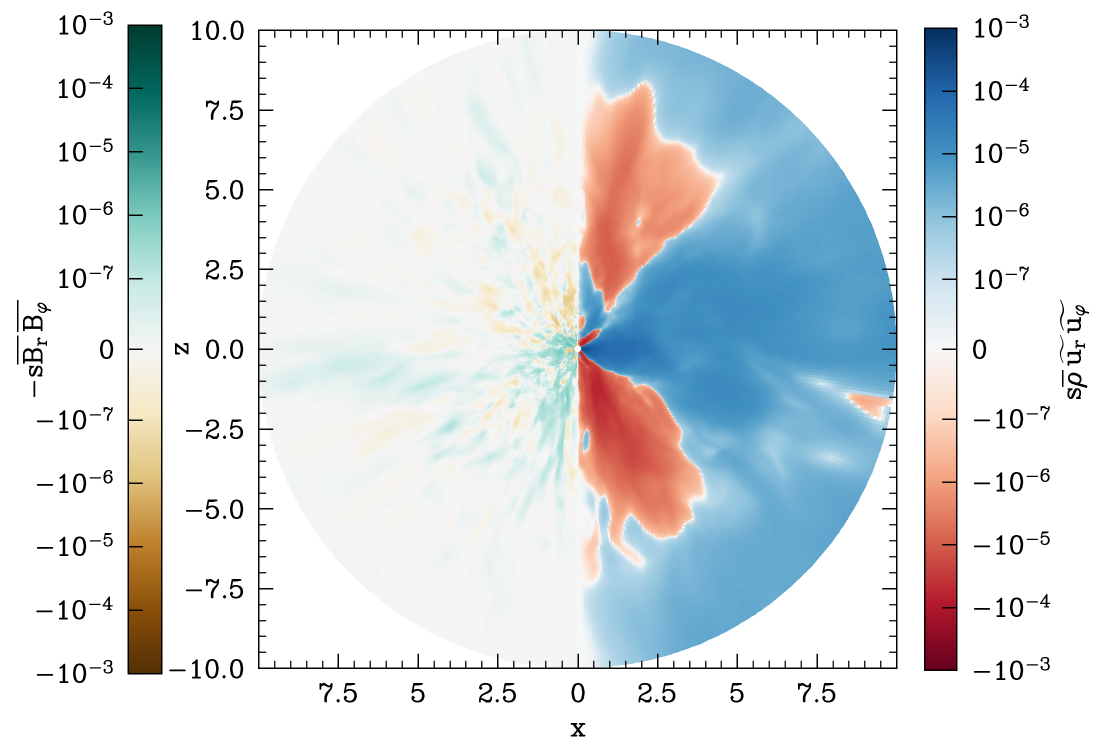} \\
              \includegraphics[ width=88mm]{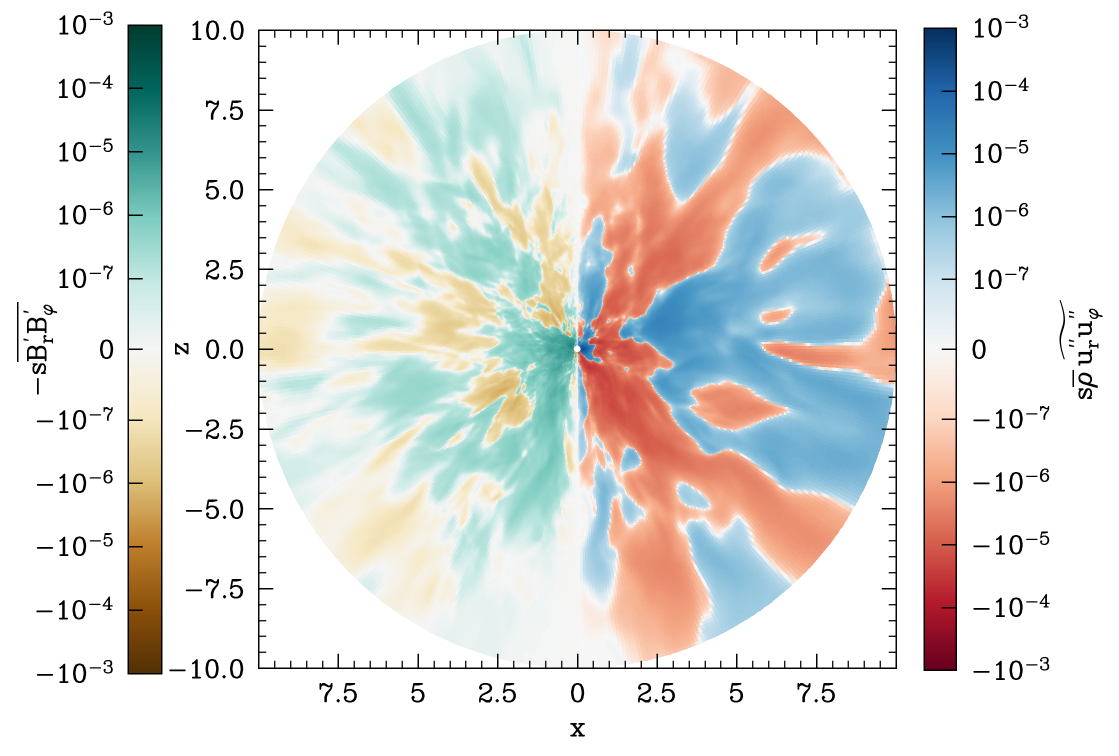} \hfill
   \caption{Same as Fig.~\ref{fig:AM_trans} but for simulation B'.}
\label{fig:AM_trans_ref}
\end{figure}

Because the time-periodic gravitational body force exerted by the binary system on its surrounding envelope is zero at the poles and maximum in the orbital plane, the dynamics of common envelopes is highly dependent on latitude. It is therefore essential to analyze the latitudinal dependence of angular momentum transport. To do so, we first show in Figs.~\ref{fig:AM_trans_full} and \ref{fig:AM_trans_full_ref} the total angular momentum radial advective flux $F_\text{adv} = s \mathcal{W}_{r\varphi}$ (see Eq.~\ref{eq:AMcons}). We see that in simulations B and B' and as in the nonmagnetic simulations of \cite{Gagnier2023}, the angular momentum advective flux is directed outward in an equatorial disk-like structure. Above and below such disk, the angular momentum advective flux is directed inward. 
We assess the relative significance and latitudinal variation of each component of the total angular momentum radial advective flux in Figs.~\ref{fig:AM_trans} and \ref{fig:AM_trans_ref}. The top panel shows that the morphology of the angular momentum radial flux is dominated by the nonmagnetic component. The radial transport of angular momentum by magnetic fields is directed outward at all latitudes and radii for simulation B. Consequently, it enhances outward transport in the equatorial disk-like structure, and damps the inward transport elsewhere. The morphology of magnetic fields driving radial transport of angular momentum is slightly more complex in simulation B' as its direction and amplitude vary with $r$ and $\theta$. From the second and third panels of Figs.~\ref{fig:AM_trans} and \ref{fig:AM_trans_ref} showing the mean and turbulent components of the total flux, we see that the turbulent transport associated with the Reynolds stress is less efficient than transport by the mean flow. This result is similar to the  nonmagnetic simulations of \cite{Gagnier2023}. Still, the complex morphology and nonnegligible amplitude of the turbulent transport may significantly damp or enhance the angular momentum radial transport  locally and may even reverse the latitudinally integrated angular momentum radial transport. We see that the magnetic radial transport of angular momentum is dominated by the turbulent Maxwell stress while the transport by mean (axisymmetric) field is negligible. 



\subsection{The value of the disk $\alpha$ parameter}
\label{sec:alpha}

\begin{figure} 
\centering
      \includegraphics[ width=88mm]{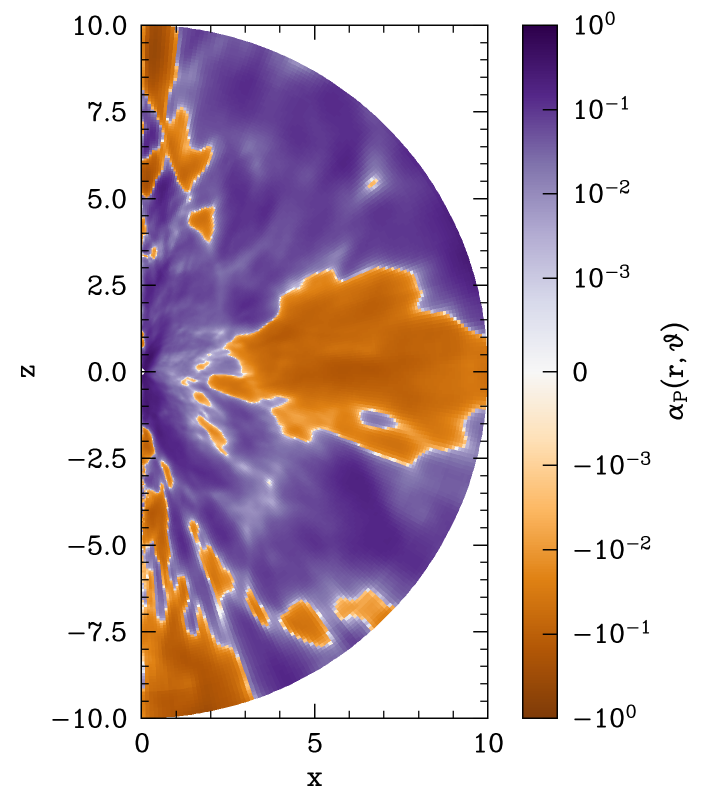} \hfill
   \caption{Reynolds-averaged turbulent stress normalized to gas pressure, $\alpha_P$, averaged over twenty orbital periods for $140 \leq t/P_{\rm orb} \leq 160$ for simulation B.}
\label{fig:alphaP_mer}
\end{figure}
\begin{figure} 
\centering
      \includegraphics[ width=88mm]{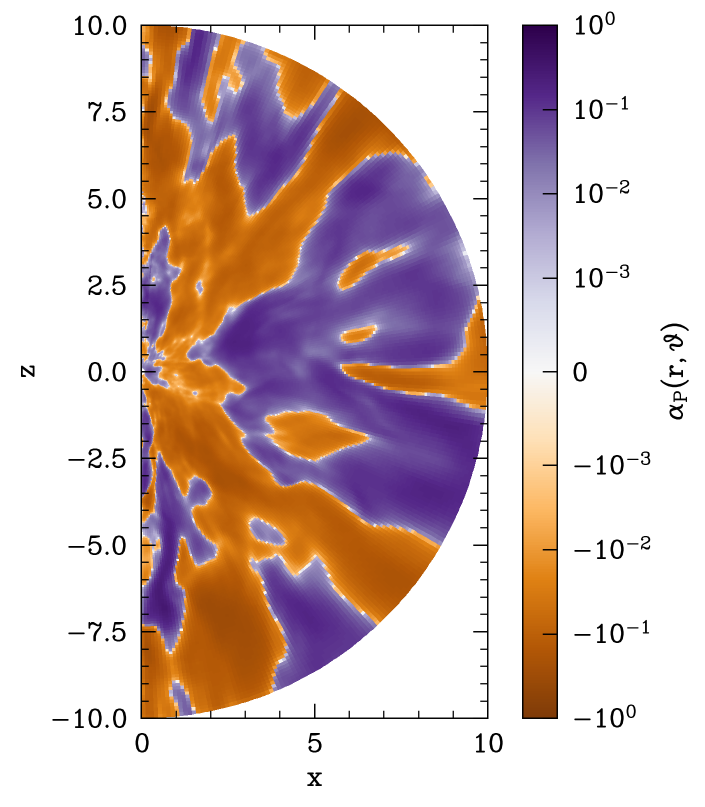} \hfill
   \caption{Same as Fig.~\ref{fig:alphaP_mer} but for simulation B'.}
\label{fig:alphaP_mer_ref}
\end{figure}

Recently, \cite{Tuna2023} have developed an $\alpha$-disk model for the long-term evolution of post-common envelope circumbinary disks. To check their assumptions and to facilitate similar studies in the future, it is important to constrain the value of $\alpha$  more accurately. In our analysis, presented in Sect.~\ref{sec:QSS} and summarized in Table~\ref{tab:runs}, we measured the volume-averaged turbulent stress normalized to the gas pressure, $\alpha_P$, as well as the normalized $r\varphi$ component of the Reynolds and Maxwell stresses. These measurements indicate that at QSS, angular momentum is globally transported outward when accretion is not prevented. The values we obtain are in agreement with previous global and local magnetohydrodynamic simulations of accretion disks \citep[e.g.,][]{Simon2009,Hawley2013,Parkin2013}. 

However, while many simulations of circumbinary disks are dominated by MRI with $\alpha_P$ mainly driven by the Maxwell stress that is predominantly positive throughout the disk \citep[e.g.,][]{Shi2012,Shi2015}, our simulations reveal a contrasting behavior. Magnetic fields in our accreting simulations are weak compared to the strength of the mean flow driven by the binary and have consequently limited direct dynamical impact \citep[in agreement with][]{Ondratschek2022}. Yet, as we have seen in Sect.~\ref{sec:ME}, magnetic fields have a considerable effect on the envelope's density structure and therefore indirectly on angular momentum transport through the mean flow and through Reynolds stress.
The Reynolds stress is the main contributor to the local turbulent flux of angular momentum. However, because it exhibits significant variations in sign and amplitude throughout the shared envelope \citep[see also][]{Gagnier2023}, its contribution to $\alpha_P$ is of the same order as that of the turbulent Maxwell stress. The turbulent Maxwell stress enhances the outward and damps the inward turbulent transport of angular momentum without changing its global morphology.
We obtain similar results when accretion through the inner boundary is prevented by reflecting boundary conditions (Figs.~\ref{fig:AM_trans_full_ref}, \ref{fig:AM_trans_ref}). Angular momentum transport is dominated by the mean flow and consists of an outward transport in a disk-like structure and an inward transport along the polar axis. The main difference comes from the more complex morphology of the Maxwell and Reynolds stress and their overall weaker amplitude resulting in near-zero net radial turbulent transport of angular momentum. The global morphology of the turbulent transport of angular momentum is rather complex as its direction varies with latitude and distance to the central binary (see Figs.~\ref{fig:alphaP_mer} and \ref{fig:alphaP_mer_ref}). Hence the relevance of employing a spatially constant value of $\alpha_P$ in post-CE circumbinary disks models may be questionable. 

Our findings also reveal that the radial transport of angular momentum is predominantly driven by the mean flow, which essentially consists of spiral density waves. This poses a significant challenge to the use of the conventional $\alpha$--viscosity formalism in 1D models, as it cannot directly account for the nondiffusive nature of angular momentum transport associated with these waves. Alternative models are needed to capture the intricate dynamics and angular momentum transport involving spiral density waves in post-CE circumbinary disks.

\subsection{Time variability of accretion}
\label{sec:accretion_variability}

\begin{figure} 
\centering
    \includegraphics[ width=88mm]{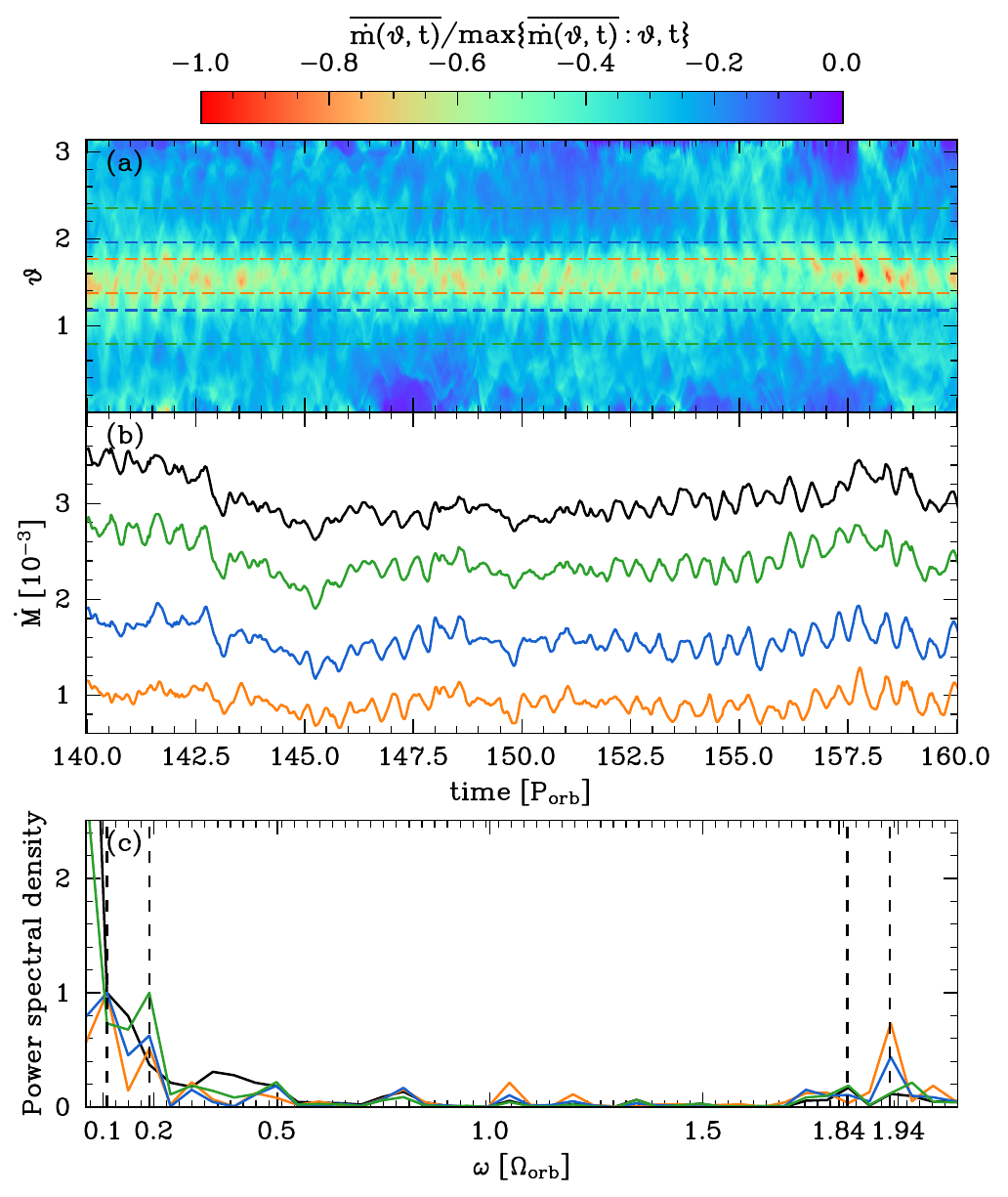} \hfill
   \caption{Detailed view on the variability of mass flux for model B. Panel (a): Space-time diagram of the local mass flux through the inner boundary. Panel (b): Time evolution of the mass accretion rate onto the binary. Panel (c): Power spectral density of the total mass accretion rate onto the binary (black line) and of the mass accretion rate in various colatitude ranges about the orbital plane (colored lines). Green lines correspond to the range $\pi/4 \le \theta \le 3\pi/4$, blue lines to the range $3\pi/8 \le \theta \le 5\pi/8$, and orange lines to the range $7\pi/16 \le \theta \le 9\pi/16$. }
\label{fig:fft}
\end{figure}

In Fig.~\ref{fig:fft}, we show a space-time diagram of the mass flux crossing the inner boundary normalized by its maximum value in the time interval $140 \leq t/P_\text{orb} \leq 160$ for our accreting model B.  As in the nonmagnetic simulations of \cite{Gagnier2023}, we see that mass accretion exhibits  periodic variability at all colatitudes. On the orbital plane, a variability frequency of $2 \Omega_\text{orb}$ is clearly visible. This frequency is associated with the quadrupolar moment contribution to the binary potential. In addition to this expected variability frequency, \cite{Gagnier2023} further found an additional low-frequency modulation of accretion $\omega \simeq 0.2\Omega_\text{orb}$, which they attribute to the presence of an eccentric and tilted overdensity, successively feeding the individual binary components through accretion streams. Remarkably, such mass accretion variability frequency are identical to those measured in circumbinary disks simulations where it is attributed to the orbiting frequency of a nonaxisymmetric overdensity (or ``lump'').
We show the power spectral density of the total mass accretion rate $\dot{M}$, as well as the mass accretion rates measured in various opening angles about the orbital plane in Fig.~\ref{fig:fft}, panel c. We find the $0.2\Omega_\text{orb}$ frequency is present, suggesting that magnetic fields do not prevent the formation of the overdensity observed in the nonmagnetic simulations of \cite{Gagnier2023}. It is however likely that higher magnetization would weaken the lump and its induced low-frequency accretion modulation \citep[][]{Noble2021}. In addition to the $0.2\Omega_\text{orb}$ frequency, we find a clear and dominating peak at $\omega \simeq 0.1\Omega_\text{orb}$. We investigate its origin in Sect.~\ref{sec:lump}. We note that the  $0.1\Omega_\text{orb}$ and  $0.2\Omega_\text{orb}$ frequencies are not visible in the power spectral density of the total mass accretion rate because of the asynchronicity of mass accretion between colatitudes. These results suggest that local analyses are necessary when studying short-term variability of accretion in CEE  \citep[][]{Gagnier2023}.

\subsection{The lump}\label{sec:lump}

 In the context of circumbinary disks, the appearance of a lump is commonly thought to be the result of the interaction between gas streams inside the cavity and the cavity edge \citep[e.g.,][]{Shi2012,Noble2012}. However, recent work by \cite{MignonRisse2023} and \cite{Rabago2023} suggest an alternative lump formation channel consisting of the merging of  Rossby Wave Instability \citep[RWI,][]{Lovelace1999,Li2000} induced large-scale vortices into a single lump. \cite{Cimerman2023} further show that such large-scale vortices likely also play a major role in angular momentum transport in CBDs by launching vortex-driven spiral density waves \citep[see also][]{Huang2019}. In circumbinary disks, RWI can be triggered from an axisymmetric bump in the inverse vortensity associated with the strong radial gradient of density at the cavity edge. In common envelopes however, there is likely no cavity, and such strong vortensity (or magnetized vortensity) maxima may not naturally emerge. Still, Rossby wave instability can be triggered by the vortensity production caused by the nonlinear damping of spiral density waves \citep{Coleman2022}. 
 Here, we investigate the existence of nonaxisymmetric overdensities in our simulations and we explore their behavior and origin. 
 
\begin{figure} 
\centering
    \includegraphics[ width=88mm]{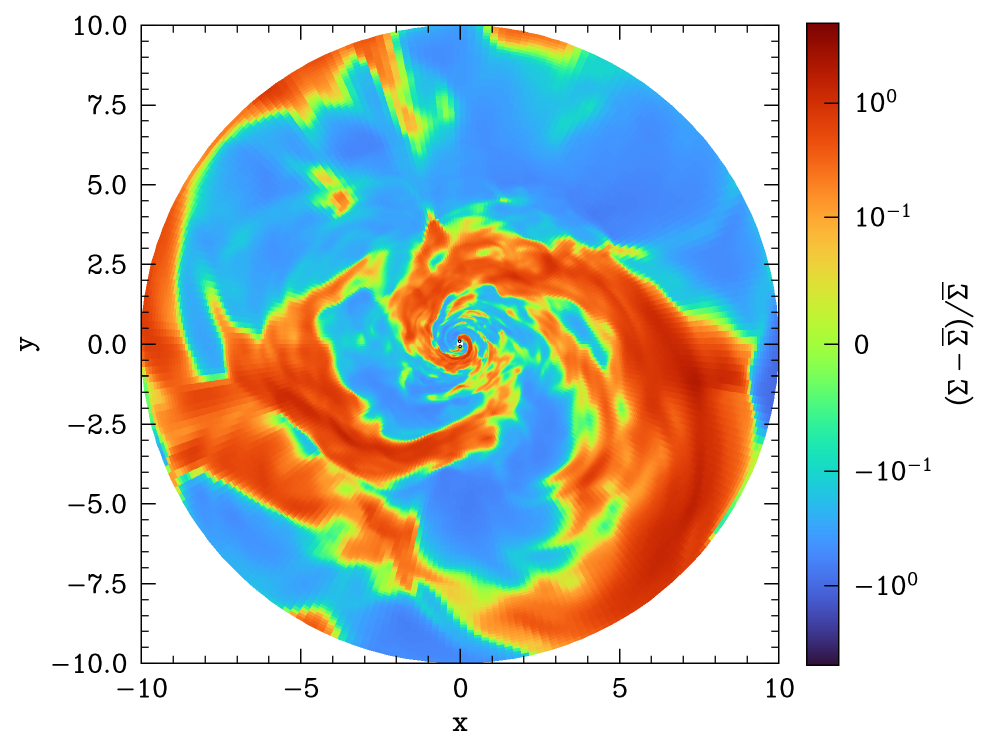} \hfill
   \caption{Snapshot of relative surface density perturbation $(\Sigma - \overline{\Sigma})/\overline{\Sigma}$ in the xy plane, at $t= 140 P_\text{orb}$.}
\label{fig:sigma}
\end{figure}

\begin{figure} 
\centering
    \includegraphics[ width=88mm]{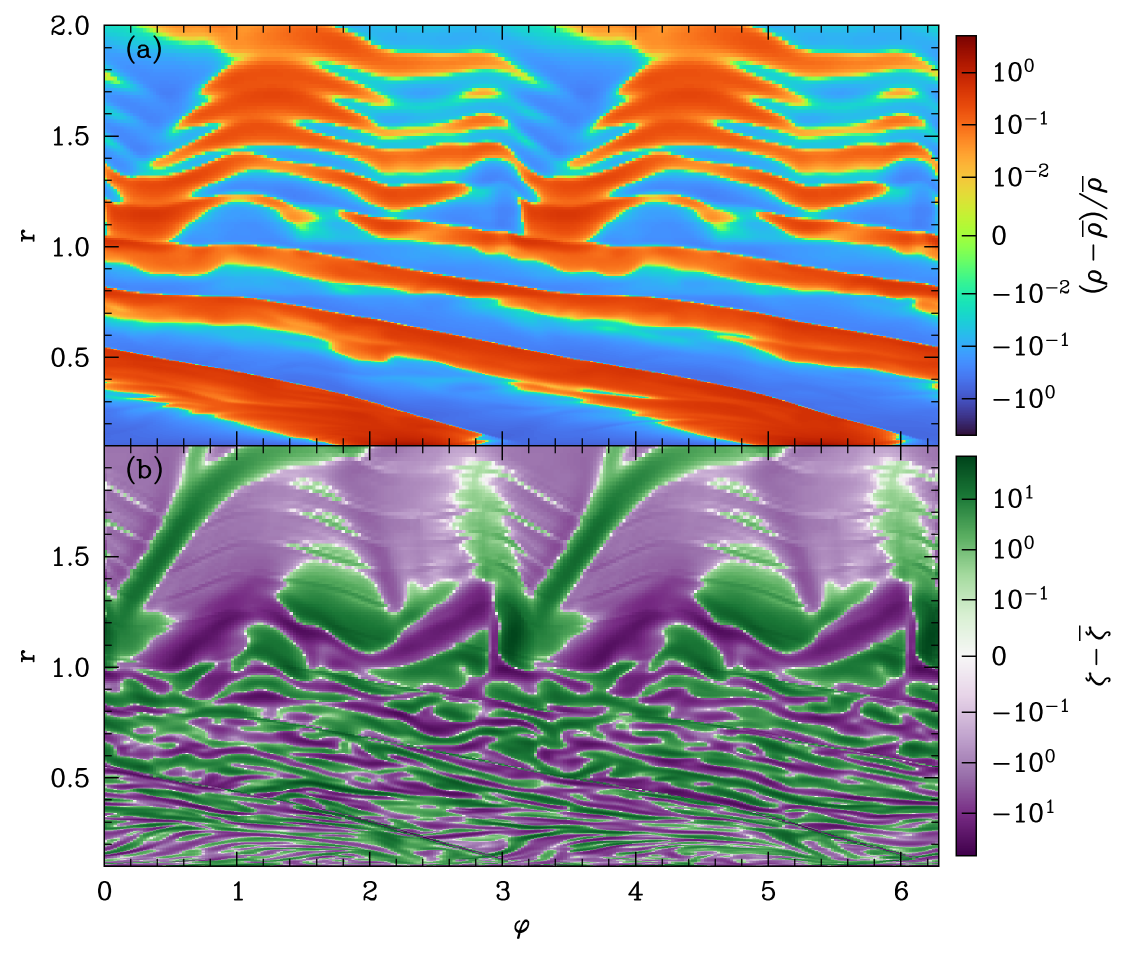} \hfill
   \caption{Snapshot of density perturbation $(\rho - \overline{\rho})/\overline{\rho}$ (panel a) and of vortensity deviation $\zeta - \overline{\zeta}$, where $\zeta = \rho^{-1} \bnabla \times \buu$, in the $r\varphi$ plane at $t= 25 P_\text{orb}$. }
\label{fig:sigmavort1}
\end{figure}

 In Fig.~\ref{fig:sigma}, we show a snapshot of the relative surface density perturbation  $(\Sigma - \overline{\Sigma})/\overline{\Sigma}$ at $t= 140 P_\text{orb}$ for simulation run B. The surface density is defined as
\begin{equation}\label{eq:sigma}
    \Sigma = \int_{7\pi/16}^{9 \pi/16} \rho r \sin \theta \dd\theta \ .
\end{equation}
We find strong and structured nonaxisymmetric overdensities taking the form of spiral density waves with a large pitch angle. The presence of a single well-defined spiral arm reaching the binary in Fig.~\ref{fig:sigma} suggests that $m=1$ modes also play a major role in the shaping of overdensities.
In Fig.~\ref{fig:sigmavort1}, we investigate the origin of such overdensities by first showing the nonaxisymmetric density perturbation on the orbital plane as well as the vortensity perturbation early in our simulation at $t= 25 P_\text{orb}$. In the inner part of the envelope, we find small scale turbulence to be fully developed, which is likely a consequence of stratified shear instability emerging from the interaction between spiral density waves generated by the binary's torque and the background velocity shear. We further note the presence of four large anticyclonic vortices located at $r \simeq 1$. These vortices likely emerge from Rossby wave instability, triggered by the vortensity production at this radius due to the nonlinear damping of spiral density waves \cite{Coleman2022}. The propagation of spiral density wavefronts through such vortices  induces their deformation, as is evident in panel b. This deformation of the wavefront entails the emergence of crests where the deformed density waves converge. Notably, two out of the four vortices exhibit sufficient strength to cause the associated wavefront crests to give rise to the formation of persistent, slowly orbiting overdensities taking the form of $m=2$ spiral density waves with large pitch angle. The existence of such overdensities prevent the spiral density waves generated by the binary's torque from propagating freely outward, resulting in material accumulation and the formation of a nonaxisymmetric lump \citep[see][for more details]{Gagnier2023}.

\begin{figure} 
\centering
    \includegraphics[ width=88mm]{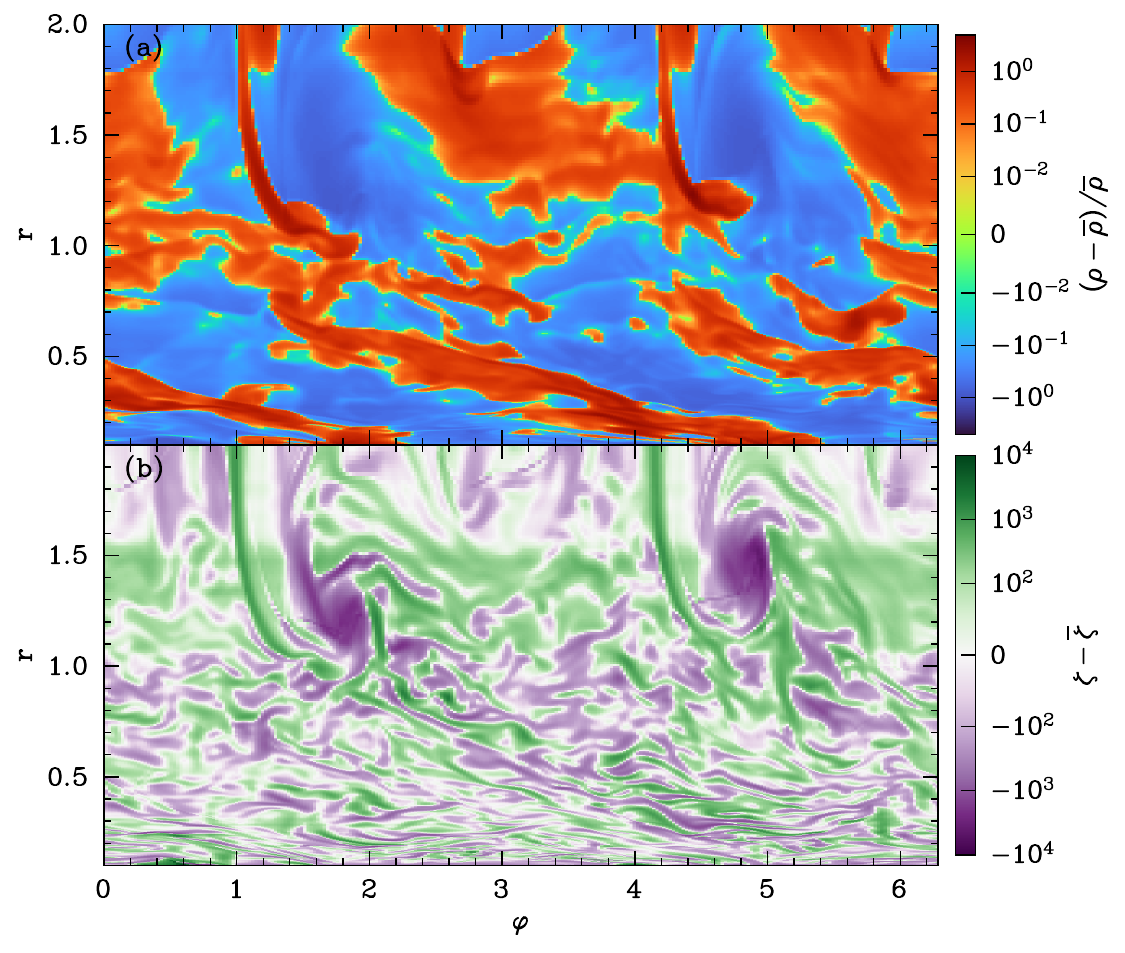} \hfill
   \caption{Same as Fig.~\ref{fig:sigmavort1} but at $t= 45 P_\text{orb}$. }
\label{fig:sigmavort2}
\end{figure}

\begin{figure} 
 \centering
     \includegraphics[ width=88mm]{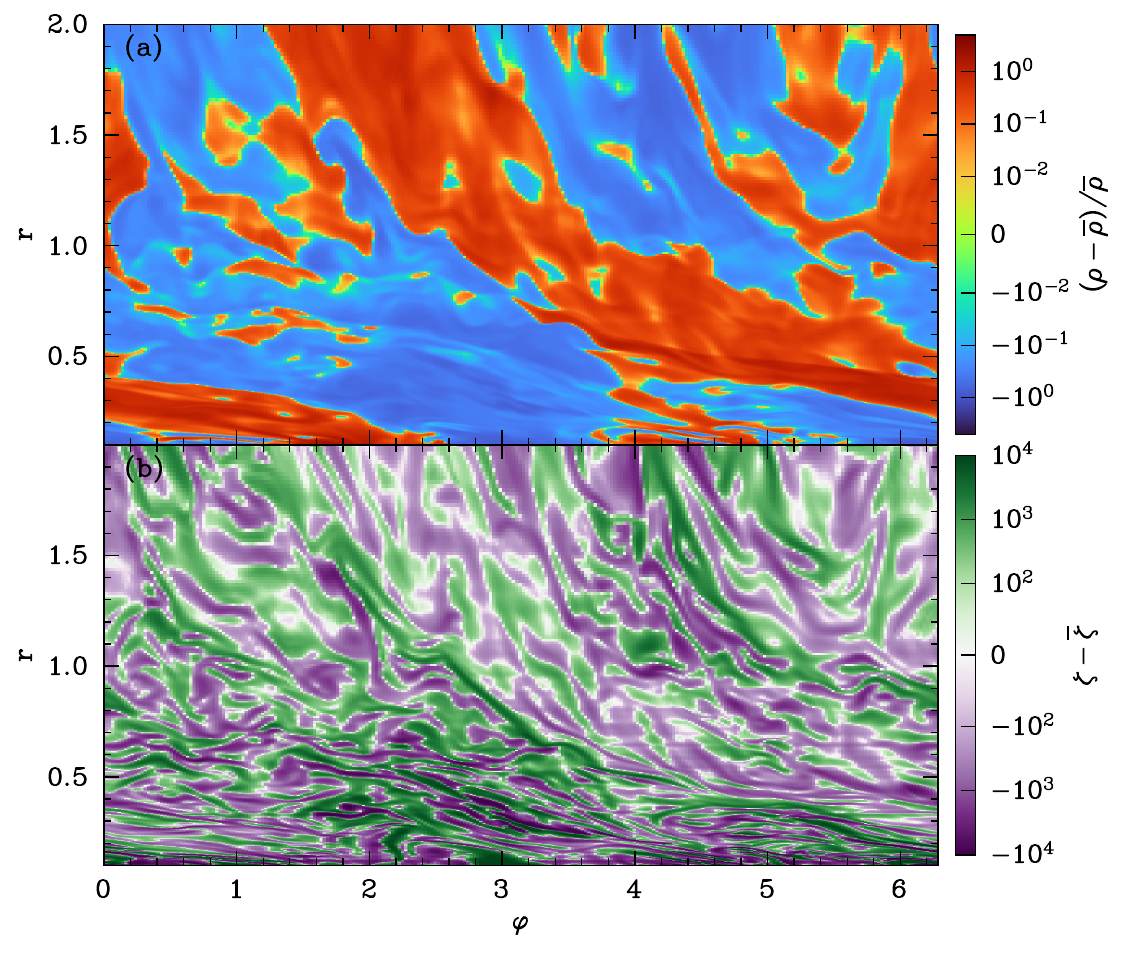} \hfill
    \caption{Same as Figs.~\ref{fig:sigmavort1} and \ref{fig:sigmavort2} but at $t= 140 P_\text{orb}$. }
\label{fig:sigmavort3}
 \end{figure}

 Eventually, the inner parts of the two spiral arms become sufficiently dense for the gravitational force exerted on them by the central binary to overcome the pressure gradient driving the envelope's expansion, leading to their inward migration. 
 In Fig.~\ref{fig:sigmavort2}, we see that the two overdensities exhibit slightly different inward migration speeds due to the imperfect mass distribution symmetry. Such inward migration of overdensities within the expanding envelope induces strong and unstable horizontal shear, giving rise to two prominent vortices. As the faster-moving overdensity approaches the binary, it experiences substantial torque and ultimately reaches the inner boundary as a singular accretion stream. This event locally disrupts the $m=2$ symmetry, favoring instead $m=1$ symmetry in the inner part of the envelope (Fig.~\ref{fig:sigmavort3}). Over time, the density of this single spiral arm diminishes as material is accreted by the binary and is eventually destroyed by the $m=2$ spiral density waves and by tidal disruption of the overdense stream, thereby restoring the $m=2$ symmetry in the binary's vicinity. This cycle repeats every $\sim 10P_\text{orb}$ and it therefore associated with the $\omega \simeq 0.1 \Omega_\text{orb}$ frequency observed in Fig.~\ref{fig:fft}. We note that this process and thus this characteristic frequency were not present in \cite{Gagnier2023}'s nonmagnetic runs. 
In simulation B, the formation and persistence of $m=1$ accretion streams can be attributed to the stabilizing effect of the toroidal magnetic fields preventing the destruction of overdensities through magnetic tension, allowing these structures to progressively increase in density. As these overdensities become denser, their gravitational attraction to the inner binary system grows stronger, leading to their inward migration. Their relatively high density enables them to reach the inner boundary and feed the binary before being destroyed.

\section{Discussions and conclusions}\label{sec:conclusions}

In this paper, we have significantly extended our previous work in \citet{Gagnier2023} on the post-dynamical inspiral phase of CEE by including magnetic fields in 3D simulations. 
Our first aim was to determine the sources of magnetic energy amplification, and the relative importance of magnetic to kinetic energy reservoir size.
We found that magnetic energy amplification arises primarily from the stretching, folding, and winding of the initial weak poloidal field due to differential rotation and turbulence. As the magnetic field strengthens, it stabilizes fluid motion, favoring azimuthal kinetic energy over radial kinetic energy during saturation. Magnetic fields significantly impact the envelope's structure and dynamics, with the magnetic energy reaching levels similar to the previous MHD simulations of \cite{Ohlmann2016b} and \cite{Ondratschek2022}, but with a much lower kinetic-to-magnetic energy ratio. Energy spectra show kinetic energy dominance at all scales, but decreasing towards smaller scales. Our analysis identifies stretching of magnetic field lines by velocity shear as the dominant source of magnetic energy, contributing 95\% during quasi-steady state with the remaining 5\% resulting from compression against magnetic pressure. 

Our second aim was to determine how kinetic and magnetic energy reservoirs are interconnected and what contributes to their evolution during the post-dynamical phase of CEE. We first decomposed the kinetic energy evolution equation into mean and turbulent contributions to determine the relative contribution of the different processes. We found the impact of magnetic fields on the mean kinetic energy evolution  to be negligible, and thus not to directly affect the mean flow. However, the interaction between turbulent shear and Maxwell stress leads to a nonnegligible sink of turbulent kinetic energy. 
We then employed energy transfer analysis to investigate the scales at which energy is transferred between kinetic and magnetic energy reservoirs. This analysis revealed that magnetic energy production by the stretching of magnetic field lines occurs at all scales and that the advection of magnetic energy within the magnetic energy reservoir indicates a forward cascade from larger to smaller scales. Notably, both field line stretching and magnetic advection peak at a critical scale of approximately $3 a_\text{b}$, that is, approximately the wavelength of the binary-driven spiral density waves. Compression effects, though weaker, play a significant role in net magnetic energy production due to the balance  between stretching and advection at this scale. Our analysis also shows that the scales at which kinetic energy is lost to magnetic energy and the scales at which magnetic energy is received from kinetic energy are different. This highlights the nonlocality of interactions between magnetic and kinetic energy in spectral space. 

Our analysis indicates the absence of  magnetic energy production on large radial scales. This suggests potential challenges in funneling and shaping radial polar outflows in planetary nebulae, possibly hindering the formation of well-defined bipolar structures. However, the presence of a strong toroidal magnetic field on the orbital plane, coupled with centrifugal forces and turbulent mixing, could respectively slow down envelope equatorial expansion and channel jittery and irregular outflows near the poles, thus leading to the emergence of highly nonspherical planetary nebulae.

By comparing our results to the nonmagnetic simulations outcomes of \cite{Gagnier2023}, our third aim was to assess the impact of magnetic fields on the binary separation evolution timescale, on angular momentum transport within the shared envelope, on the short-term variability of mass accretion onto the binary, and on the formation of overdensities. We found that magnetic fields play little to no role in the binary separation evolution. This result may, however, be closely tied to our choice of boundary conditions for the magnetic field and may thus call for further investigation through numerical simulations on the scale of binary interaction that can accurately resolve the accretion flow and the interaction between magnetic fields and the cores. This will be explored in future works.

We found magnetic fields do not change the disk-like morphology of radial angular momentum transport found by \citet{Gagnier2023}. However, when accretion onto the central binary is allowed, we found that turbulent Maxwell stress has a comparable or even larger net contribution to the radial angular momentum transport than the Reynolds stress.  Turbulent Maxwell stress transport of angular momentum points outward at all radii and it locally reverses the direction of angular momentum transport that is otherwise dominated by the mean-flow advection. When accretion is prevented by imposing reflecting boundary conditions, the net contribution of Maxwell stress to the radial transport of angular momentum is negligible compared to the contribution from Reynolds stress and mean-flow.

Our investigation of the intricate dynamics of nonaxisymmetric overdensities within the common envelope of binary star systems reveals their pivotal role in modulating mass accretion onto the binary components. 
Similarly to \cite{Gagnier2023}, we find mass accretion to exhibit distinct frequencies, including the $2 \Omega_\text{orb}$ frequency stemming from the quadrupolar moment of the binary's potential and a lower-frequency $\simeq 0.2\Omega_\text{orb}$ modulation associated with overdensities known as ``lumps'' in the context of circumbinary disks.
Our analysis uncovers the origins of these various overdensities: they form as a result of the intricate interplay between spiral density waves launched by the central binary and spiral density waves with a much larger pitch angle, which arise from interaction with vortices. Additionally, we unveiled the emergence of $m=1$ accretion streams associated with the $\simeq 0.1\Omega_\text{orb}$ frequency. These streams owe their existence to the stabilizing effect of the magnetic tension from the strong toroidal field about the orbital plane, which prevents overdensities from being destroyed by turbulence and enables them to accumulate mass and eventually migrate towards the binary. Ultimately, these denser structures are destroyed by spiral waves and tidal forces in the binary's vicinity.

Finally, Figs.~\ref{fig:snap925} and \ref{fig:snap925_B} show a feature that looks like an ``S-shaped'' cavity. We saw this cavity already in our previous work \citep{Gagnier2023}. This cavity results from the interplay of centrifugal force, which ``pushes'' material perpendicularly away from the rotation axis, and stochastic convection motion disrupting the symmetry of the resulting low-density chimney. Although our simulations do not exhibit the launching of jets or jet-like outflows, the presence of the cavity suggests the possibility of collimation of a ``wobbling'' jet, which could originate around one of the cores. It is exciting to speculate that ``wobbling'' of jets seen in many planetary nebulae might not arise due to the precession from the second core, but instead due to time-variable S-shaped cavity in the circumbinary envelope. Resolving the inner region, specifically inside the orbit, is probably necessary to determine whether outflows are actually launched. We plan to study this aspect in our future work.

This study has provided a comprehensive examination of the impact of magnetic fields on the post-dynamical inspiral phase of common envelope evolution. We have revealed the intricate interplay between magnetic energy amplification, energy reservoirs, and the dynamics of the envelope, shedding light on the mechanisms responsible for magnetic energy amplification and transfer. Our findings have also elucidated the role of magnetic fields in angular momentum transport and the formation of nonaxisymmetric overdensities within the common envelope. However, further investigation is needed to fully comprehend the role of magnetic fields in this context and their wider implications. In particular, recent circumbinary disk simulations have showed that resolving the cavity within the computational domain is essential for accurately measuring torques, orbital evolution timescales, and orbital eccentricity excitation \citep[e.g.,][]{Tang2017,Moody2019,Munoz2019,Munoz2020,Duffell2020,Tiede2020,Dittmann2021,Combi2022}, and to self-consistently launch jets \citep[e.g.,][]{Gold2014}. Adopting this approach is likely even more crucial in the context of CEE because of the probable absence of a low-density cavity. Instead, we anticipate that the region inside the binary orbit exhibits highly complex gas dynamics and substantial magnetic energy amplification. We plan to explore these phenomena in future works.

\begin{acknowledgements}
We thank the anonymous referee for comments that improved this paper. We acknowledge fruitful discussions with K. Tomida, S. Takasao, G. Wafflard-Fernandez, and G. Lesur. 
The research of DG and OP has been supported by Horizon 2020 ERC Starting Grant `Cat-In-hAT' (grant agreement no. 803158).
This work was supported by the Ministry of Education, Youth and Sports of the Czech Republic through the e-INFRA CZ (ID:90254). 
\end{acknowledgements}

\bibliographystyle{aa}
\bibliography{bibnew}

\onecolumn
\begin{appendix}
\section{Horizontally integrated energy spectrum}\label{app:ES}
The Fourier transform of a function $f(\br) = f(r,\theta,\varphi)$ reads
\begin{equation}\label{eq:fourier1}
    \widehat{f}(\boldsymbol{k}) = \widehat{f}(k,\theta_k,\varphi_k) = \int_0^{2\pi} \int_0^\pi \int_0^\infty f(\br) e^{i \boldsymbol{k}\cdot \br} r^2 \sin \theta \dd r  \dd \theta \dd \varphi \ .
 \end{equation}
 We expand both $f$ and the Fourier kernel $e^{i \boldsymbol{k}\cdot \br}$ on the spherical harmonics basis
  \begin{equation}\label{eq:fylm}
f(\br) = \sum_{\ell=0}^\infty \sum_{m=-\ell}^\ell f_m^\ell (r) Y_\ell^m (\theta,\phi) \ ,
 \end{equation}
 and
 \begin{equation}\label{eq:exp}
e^{i \boldsymbol{k}\cdot \br} = 4 \pi \sum_{\ell=0}^\infty \sum_{m=-\ell}^\ell i^\ell j_\ell(kr) Y_\ell^m (\theta,\phi)^\ast Y_\ell^m (\theta_k,\phi_k) \ ,
 \end{equation}
 where  $Y_\ell^m$ is the usual scalar spherical harmonics function and $j_\ell(z)$ is the spherical Bessel function of the first kind of order $\ell$. Injecting (\ref{eq:fylm}) and (\ref{eq:exp}) into (\ref{eq:fourier1}) yields
\begin{equation}\label{eq:fourier}
  \widehat{f}(\boldsymbol{k})  =    \sum_{\ell=0}^\infty \sum_{m=-\ell}^\ell F_\ell^m(k) Y_\ell^m (\theta_k,\phi_k) \ ,
\end{equation}
 where $F_\ell^m(k)$ is the $\ell$\textsuperscript{th} order spherical Bessel transform of the spherical harmonics expansion coefficients
 \begin{equation}\label{eq:Hankel}
     F_\ell^m(k) =4 \pi  i^\ell \int_0^\infty j_\ell(kr) f_m^\ell (r) r^2 \dd r \ .
 \end{equation}
 The  horizontally integrated (in Fourier space) energy spectrum of $f(\br)$ finally reads
 \begin{equation}\label{app:Ef_k}
\begin{aligned}
    E_f(k) &= k^2  \int_0^{2\pi} \int_0^\pi  \widehat{f}(\boldsymbol{k})  \widehat{f}(\boldsymbol{k})^\ast  \sin \theta_k   \dd \theta_k \dd \varphi_k \\ &=  k^2 \sum_{\ell=0}^\infty \sum_{m=-\ell}^\ell F_\ell^m(k)  F_\ell^m(k)^\ast      \ .
    \end{aligned}
 \end{equation}
 This quantity relates to the total energy through Parseval's theorem
\begin{equation}
\langle E_f \rangle= \int_{\partial V} |f(\br)|^2 \dd V = \frac{1}{(2\pi)^3}\int_0^\infty E_f(k)\dd k \ .
\end{equation}
\section{Magnetic energy evolution equation}\label{app:EB}
The magnetic energy evolution equation (\ref{eq:Eb_ev}) is derived from the dot product of $\bB$ with the induction equation (\ref{eq:induc}), 
that is
\begin{equation}
    \frac{1}{2}\frac{\partial \left(\bB \cdot \bB \right)}{\partial t} - \bB \cdot \left(\bB \cdot \bnabla \buu \right) + \bB \cdot \bB (\bnabla \cdot \buu) + \bB \cdot \left( \buu \cdot \bnabla \bB \right) = 0 \ .
\end{equation}
Using the fact that $\bB \cdot \left( \buu \cdot \bnabla \bB \right) = \buu \cdot \left( \bnabla \bB \cdot \bB \right) $ and $\bnabla \bB \cdot \bB = \bB \cdot \bnabla \bB + \bB \times \left(\bnabla \times \bB \right)$, yields
\begin{equation}
    \frac{1}{2}\frac{\partial \left(\bB \cdot \bB \right)}{\partial t} - \bB \cdot \left(\bB \cdot \bnabla \buu \right) + E_B \bnabla \cdot \buu + \bnabla \cdot \left( E_B \buu \right) = 0 \ , 
\end{equation}
which finally yields the magnetic energy evolution equation (\ref{eq:Eb_ev})
\begin{equation}
    \frac{1}{2}\frac{\partial \left(\bB \cdot \bB \right)}{\partial t} - \left(\bB \otimes \bB \right) {:} \bnabla \buu + E_B \bnabla \cdot \buu + \bnabla \cdot \left( E_B \buu \right) = 0 \ . 
\end{equation}
\section{Mean and turbulent kinetic energy evolution equations}\label{app:MTKE}

The mean kinetic energy evolution equation (\ref{eq:Ekmean_rt}) is derived from the Reynolds averaged momentum and mass conservation equations
\begin{align}\label{eq:B1}
 &   \frac{\partial \overline{\rho} \widetilde{\buu}}{\partial t} + \bnabla \cdot (\overline{\rho}  \but \otimes \but) + \bnabla \cdot (  \overline{\rho \bup\otimes \bup}) = - \bnabla \overline{P}  +  \bnabla \cdot \overline{\bsigma}-\overline{\rho \bnabla \Phi}+ \bnabla \cdot \overline{\btau} \ ,\\ 
 &   \frac{\partial \overline{\rho}}{\partial t} + \bnabla \cdot (\overline{\rho} \but) = 0 \ . \label{eq:B2}
\end{align}
Taking the dot product of $-\but$  with  Eq.~(\ref{eq:B2}) and adding it to Eq.~(\ref{eq:B1}), making use of the divergence of a dyad formula $\bnabla \cdot( \boldsymbol{a} \otimes \boldsymbol{b}) = (\bnabla \cdot \boldsymbol{a})\boldsymbol{b} + \boldsymbol{a} \cdot \bnabla \boldsymbol{b}$, yields
\begin{equation}\label{eq:B3}
    \overline{\rho}  \frac{\partial \but}{\partial t} +  (\overline{\rho}  \but \cdot \bnabla) \but + \bnabla \cdot (\overline{\rho \bup\otimes \bup}) = - \bnabla \overline{P}  +  \bnabla \cdot \overline{\bsigma}-\overline{\rho \bnabla \Phi}+ \bnabla \cdot \overline{\btau}  \ .
\end{equation}
We then take the dot product of $\but$ with Eq.~(\ref{eq:B3}) by and we multiply Eq.~(\ref{eq:B2}) by $(\but \cdot \but)/2$, respectively yielding
\begin{equation}\label{eq:B4a}
     \frac{1}{2} \overline{\rho}\frac{\partial (\but \cdot \but)}{\partial t}  + \overline{\rho} \but \cdot \bnabla \frac{(\but \cdot \but)}{2} + \but \cdot \left(\bnabla \cdot (\overline{\rho \bup\otimes \bup} )\right)= - \but \cdot \bnabla \overline{P}  +  \but \cdot \left(\bnabla \cdot \overline{\bsigma}\right)- \overline{\rho \but \cdot \bnabla \Phi}+ \but \cdot \left(\bnabla \cdot \overline{\btau}\right) \ , 
\end{equation}
and
\begin{equation} \label{eq:B4b}
   \frac{(\but \cdot \but)}{2}\frac{\partial \overline{\rho}}{\partial t} + \bnabla \cdot \left(\overline{\rho} \but \frac{(\but \cdot \but)}{2} \right) - \overline{\rho} \but \cdot \bnabla \frac{(\but \cdot \but)}{2} = 0 \ ,
\end{equation}      
where we have used the vector identity $\bnabla (\boldsymbol{a} \cdot \boldsymbol{a})/2 = (\boldsymbol{a} \cdot \bnabla)\boldsymbol{a} + \boldsymbol{a} \times (\bnabla \times \boldsymbol{a})$. 
Finally, we add Eqs.~(\ref{eq:B4a}) and (\ref{eq:B4b}) together to obtain the Reynolds averaged mean kinetic energy evolution equation (\ref{eq:Ekmean_rt})
\begin{equation}
    \frac{1}{2}\frac{\partial \overline{\rho}(\but \cdot \but)}{\partial t} + \bnabla \cdot \left(\overline{\rho}\but \frac{(\but \cdot \but)}{2} \right)+   \bnabla \cdot \left(\overline{\rho}  \widetilde{\bup\otimes \bup} \cdot \but \right) = \overline{\rho}  \widetilde{\bup\otimes \bup} {:} \bnabla \but - \but \cdot \bnabla \overline{P}  +  \but \cdot \left(\bnabla \cdot \overline{\bsigma}\right)-  \but\overline{ \cdot \rho \bnabla \Phi}+ \but \cdot \left(\bnabla \cdot \overline{\btau}\right)  \ .
\end{equation}
Here, we have used the identity $\bnabla \cdot (\boldsymbol{\rm T} \cdot \boldsymbol{a}) = \boldsymbol{\rm T} {:} \bnabla \boldsymbol{a} + (\bnabla \cdot \boldsymbol{\rm T} ) \cdot \boldsymbol{a}$, where $\boldsymbol{\rm T}$ is a rank 2 tensor. The colon symbol indicates a Frobenius inner product $ \boldsymbol{\rm T} {:} \boldsymbol{\rm G} =  {\rm T}_{ij}{\rm G}_{ij} = {\rm Tr}(\boldsymbol{\rm T} \boldsymbol{\rm G}^\intercal)$.
To obtain the turbulent kinetic evolution equation (\ref{eq:Ekturb_rt}), we first expand the differentiated scalar and vector fields in the convective form of Eq.~(\ref{eq:mom}), except for the directional derivative $(\buu \cdot \bnabla)$, and we take the dot product of $\bup$ with the result. This yields   
\begin{equation}\label{eq:B5}
     \frac{1}{2}\rho\frac{\partial(\bup \cdot \bup)}{\partial t} +  \rho \bup \cdot \frac{\partial \but}{\partial t} + \rho \buu \cdot \bnabla \frac{(\bup \cdot \bup)}{2}+ \rho  \buu\otimes \bup {:} \bnabla \but  = - \bup \cdot \bnabla \left (\overline{P} + P^\prime \right)  +  \bup \cdot \left(\bnabla \cdot \bsigma\right)-  \rho \bup  \rho \bnabla \left(\Phi + \Phi'\right) +  \bup \cdot \left(\bnabla \cdot \btau\right)  \ .
\end{equation}
Here, we have used the same identities as for the derivation of the mean kinetic energy equation. We then multiply Eq.~(\ref{eq:mass}) by $(\bup \cdot  \bup)/2$ and we add the result to Eq.~(\ref{eq:B5}) to obtain
\begin{equation}\label{eq:B6}
   \frac{1}{2}  \frac{\partial \rho(\bup \cdot \bup)}{\partial t} +  \rho \bup \cdot \frac{\partial \but}{\partial t} +  \bnabla \cdot \left( \rho \buu \frac{(\bup \cdot \bup)}{2} \right)+ \rho  \buu\otimes \bup {:} \bnabla \but  = - \bup \cdot \bnabla \left (\overline{P} + P^\prime \right)  +  \bup \cdot \left(\bnabla \cdot \bsigma\right)-  \rho \bup  \rho \bnabla \left(\Phi + \Phi'\right) +  \bup \cdot \left(\bnabla \cdot \btau\right)  \ .
\end{equation}
Finally, taking the Reynolds average of Eq.~(\ref{eq:B6}) and making use of the Favre average properties
\begin{align}
\overline{\rho \bup} &= 0 \ ,\\
\widetilde{\bup} &= 0 \ ,\\ 
\overline{\rho \buu \otimes \buu } &= \overline{\rho}\but \otimes \but + \overline{\rho \bup \otimes \bup }  \ ,  
\end{align}
yields the turbulent kinetic evolution equation (\ref{eq:Ekturb_rt})
\begin{equation}
\begin{aligned}
    &\frac{1}{2}\frac{\partial \overline{\rho}(\widetilde{\bup \cdot \bup})}{\partial t} + \bnabla \cdot \left(\overline{\rho}  \but \frac{(\widetilde{\bup \cdot \bup})}{2}\right)
   - \bnabla \cdot \left( \overline{\bsigma \cdot \bup} -   \overline{\rho \bup \frac{(\bup \cdot \bup)}{2}} - \overline{P^\prime \bup} \right)=  - \overline{\rho}  \widetilde{\bup\otimes \bup} {:} \bnabla \but - \overline{\bsigma {:} \bnabla \bup} - \overline{\bup} \cdot \bnabla \overline{P} \\&+ \overline{P^\prime \bnabla \cdot \bup} - \overline{\rho \bup \cdot \bnabla \Phi^\prime} + \overline{\bup \cdot \left( \bnabla \cdot \btau \right)} \ ,
  \end{aligned}
\end{equation}
where we have expanded the pressure perturbation term into an advective term and a pressure dilatation term, and the magnetic term into an advective term and a term associated with the interaction between turbulent shear and Maxwell stress.

\section{Transfer functions}\label{app:TF}
\subsection{Magnetic energy transfer equation}
The magnetic energy transfer equation can be simply derived by taking the dot product of $\widehat{\bB}$ with the complex conjugate of the horizontally integrated Fourier transformed induction equation 
\begin{equation}
    \widehat{\partial_t \bB}^\ast + \widehat{\bnabla \cdot \left( \buu \otimes \bB\right)}^\ast -  \widehat{\bnabla \cdot \left( \bB \otimes \buu\right)}^\ast = 0 \ ,
\end{equation}
adding the conjugate of the result and dividing by two,
\begin{equation}\label{eq:magtrans}
    \dot{E}_B(k)= T_{\rm MT}(k) +  T_{\rm MP}(k) \ ,
\end{equation}
where, from Eq.~(\ref{app:Ef_k}),
\begin{equation}
    \dot{E}_B(k)=\frac{k^2}{2} \int_0^{2\pi} \int_0^\pi {\partial_t \left( \widehat{\bB} \cdot \widehat{\bB}^\ast \right)} \sin \theta_k   \dd \theta_k \dd \varphi_k \ .
\end{equation}
The first term on the right-hand side of Eq.~(\ref{eq:magtrans}) represents the rate of energy transfer from the kinetic energy reservoir to the $k$--component of the magnetic energy reservoir by the stretching of the field lines against magnetic tension force \citep[e.g.,][]{PG2010}
\begin{equation}
     T_{\rm MT}(k)  =   \frac{k^2}{2}\int_0^{2\pi} \int_0^\pi \left( \widehat{\bB} \cdot \left( \widehat{\bB \cdot \bnabla \buu}^\ast \right)+ \widehat{\bB}^\ast \cdot \left( \widehat{\bB \cdot \bnabla \buu} \right) \right) \sin \theta_k   \dd \theta_k \dd \varphi_k  \ .
\end{equation}
The second term on the right-hand side of Eq.~(\ref{eq:magtrans}) represents the rate of energy transfer to the $k$--component of the magnetic energy reservoir by advection and compression against magnetic pressure
\begin{equation}
        T_{\rm MP}(k)=   T_{\rm MC}(k) +  T_{\rm MA}(k) \ , 
\end{equation}
where
\begin{align}
    T_{\rm MC}(k) & =-\frac{k^2}{2}\int_0^{2\pi} \int_0^\pi \left( \widehat{\bB} \cdot \left(\widehat{\bB \bnabla \cdot \buu}\right)^\ast + \widehat{\bB}^\ast \cdot \left(\widehat{\bB \bnabla \cdot \buu}\right)  \right)\sin \theta_k   \dd \theta_k \dd \varphi_k \ , \\
    T_{\rm MA}(k) &= -\frac{k^2}{2}\int_0^{2\pi} \int_0^\pi \left(\widehat{\bB} \cdot \left(\widehat{\buu \cdot \bnabla \bB} \right)^\ast + \widehat{\bB}^\ast \cdot \left(\widehat{\buu \cdot \bnabla \bB} \right) \right) \sin \theta_k   \dd \theta_k \dd \varphi_k  \ .
\end{align}
As \cite{Rempel2014} and \cite{Grete2017}, we note that we can split $T_{\rm MC}(k)$ in two, such that the terms underlying $T_{\rm MA}(k) + 0.5 T_{\rm MC}(k)$ in the real space may be identified with a magnetic energy advective transport to other scales within the magnetic energy reservoir (i.e., magnetic cascade)
\begin{equation}
-\bB \cdot \left( \buu \cdot \bnabla \bB + \frac{\bB}{2} \bnabla \cdot \buu \right)= -\bnabla \cdot \left( \buu \frac{(\bB \cdot \bB)}{2}\right) \ . 
\end{equation}
Similarly, the terms underlying $T_{\rm MT}(k) + 0.5 T_{\rm MC}(k)$ in the real space may be identified with energy transfer from kinetic to magnetic energy reservoir via Lorentz force and the remaining nonadvective terms of the Poynting flux \citep[e.g.,][]{Rempel2014,Grete2017}, 
\begin{equation}\label{app:KBTKBC}
\bB \cdot \left( \bB \cdot \bnabla \buu - \frac{\bB}{2} \bnabla \cdot \buu \right) = \bnabla \cdot \left(\bB (\buu \cdot \bB) - \buu \frac{(\bB \cdot \bB)}{2} \right) - \buu \cdot \left( \bnabla \cdot \bsigma \right) . 
\end{equation}
The magnetic transfer functions are related to the magnetic energy rate of change in real space as follows
\begin{equation}
\begin{aligned}
    \langle \dot{E}_{B, \text{stretch}}\rangle &= \frac{1}{(2\pi)^3} \int_0^\infty T_{\rm MT}(k) \dd k \ , \\
     \langle \dot{E}_{B, \text{exp}}\rangle &= \frac{0.5}{(2\pi)^3} \int_0^\infty T_{\rm MC}(k) \dd k \ ,\\
      \langle \dot{E}_{B, \text{adv}}\rangle &= \frac{1}{(2\pi)^3} \int_0^\infty T_{\rm MA}(k) \dd k + \frac{0.5}{(2\pi)^3} \int_0^\infty T_{\rm MC}(k) \dd k \ .
\end{aligned}
\end{equation}
\subsection{Kinetic energy transfer equation}
Following \cite{Kida1990} and \cite{Grete2017}, we introduce a new variable
\begin{equation}
    \bw = \sqrt{\rho} \buu \ ,
\end{equation}
which, combining mass and momentum conservation equations follows
\begin{equation}\label{eq:wi}
      \frac{\partial \bw}{\partial t} = -\buu \cdot \bnabla \bw - \frac{1}{2}\bw \bnabla \cdot \buu + \frac{1}{\sqrt{\rho}} \bnabla \cdot \left(\bB \otimes \bB \right) - \frac{1}{\sqrt{\rho}}\bnabla P - \frac{1}{2\sqrt{\rho}} \bnabla \left( \bB \cdot \bB\right) - \sqrt{\rho}\bnabla \Phi  \ .
\end{equation}
The kinetic energy transfer equation is derived by taking the dot product of $\widehat{\bw}$ with the complex conjugate of the horizontally integrated Fourier transform of Eq.~\ref{eq:wi}, adding the conjugate of the result and dividing by two. This tields


\begin{equation}
    \dot{E}_K(k) = T_{\rm KK}(k) + T_{\rm KL}(k) + T_{\rm KP }(k) \ ,
\end{equation}
where
\begin{equation}
    \dot{E}_K(k)=\frac{k^2}{2} \int_0^{2\pi} \int_0^\pi {\partial_t \left( \widehat{\bw} \cdot \widehat{\bw}^\ast \right)} \sin \theta_k   \dd \theta_k \dd \varphi_k \ .
\end{equation}
Furthermore,
\begin{equation}
    T_{\rm KK}(k) =  T_{\rm KKa}(k) + T_{\rm KKb}(k)
\end{equation}
corresponds to a kinetic energy transport to other scales within the kinetic energy reservoir (kinetic cascade), and
\begin{align}
    T_{\rm KKa}(k) &= -\frac{k^2}{2}\int_0^{2\pi} \int_0^\pi \left( \widehat{\bw} \cdot \left( \widehat{\buu \cdot \bnabla \bw}\right)^\ast + \widehat{\bw}^\ast \cdot \left( \widehat{\buu \cdot \bnabla \bw}\right)\right) \sin \theta_k   \dd \theta_k \dd \varphi_k \\ 
     T_{\rm KKb}(k) &= -\frac{k^2}{4}\int_0^{2\pi} \int_0^\pi \left( \widehat{\bw} \cdot \left(\widehat{\bw\bnabla \cdot \buu}\right)^\ast + \widehat{\bw}^\ast \cdot \left(\widehat{\bw\bnabla \cdot \buu}\right)\right) \sin \theta_k   \dd \theta_k \dd \varphi_k \ .
\end{align}


\begin{equation}
    T_{\rm KL}(k) = T_{\rm KLa}(k) + T_{\rm KLb}(k)
\end{equation}
represents the rate of energy transfer from the magnetic energy reservoir to the k–component of the kinetic energy reservoir by the work of the Lorentz force via magnetic tension and magnetic pressure, with
\begin{align}
    T_{\rm KLa}(k) &= \frac{k^2}{2}\int_0^{2\pi} \int_0^\pi \left(  \widehat{\bw}\cdot \left(\widehat{\frac{1}{\sqrt{\rho}} \bnabla \cdot ( \bB \otimes \bB)}\right)^\ast +   \widehat{\bw}^\ast\cdot \left(\widehat{\frac{1}{\sqrt{\rho}} \bnabla \cdot ( \bB \otimes \bB)}\right)\right) \sin \theta_k   \dd \theta_k \dd \varphi_k \\ 
     T_{\rm KLb}(k) &= -\frac{k^2}{4}\int_0^{2\pi} \int_0^\pi \left( \widehat{\bw} \cdot \widehat{\frac{1}{\sqrt{\rho}} \bnabla (\bB \cdot \bB)}^\ast +\widehat{\bw}^\ast \cdot \widehat{\frac{1}{\sqrt{\rho}} \bnabla (\bB \cdot \bB)}\right) \sin \theta_k   \dd \theta_k \dd \varphi_k \ .
\end{align}
Finally, 
\begin{equation}
  T_{\rm KP }(k) =   T_{\rm KPa }(k) +   T_{\rm KPb }(k) 
\end{equation}
is the energy transfer from pressure forces and energy injection from the binary potential, with
\begin{align}
    T_{\rm KPa}(k) &= -\frac{k^2}{2}\int_0^{2\pi} \int_0^\pi \left(  \widehat{\bw} \cdot \widehat{\frac{1}{\sqrt{\rho}}\bnabla P}^\ast +  \widehat{\bw}^\ast \cdot \widehat{\frac{1}{\sqrt{\rho}}\bnabla P} \right) \sin \theta_k   \dd \theta_k \dd \varphi_k \\ 
     T_{\rm KPb}(k) &= -\frac{k^2}{2}\int_0^{2\pi} \int_0^\pi \left( \widehat{\bw} \cdot \widehat{\sqrt{\rho}\bnabla \Phi}^\ast +\widehat{\bw}^\ast\cdot\widehat{\sqrt{\rho}\bnabla\Phi}\right) \sin \theta_k   \dd \theta_k \dd \varphi_k \ .
\end{align}
\end{appendix}

\end{document}